\documentclass[12pt]{article}

\usepackage{framed}
\usepackage[english]{babel}
\usepackage[utf8x]{inputenc}
\usepackage[T1]{fontenc}
\usepackage{apacite}
\usepackage{mathrsfs}
\usepackage{listings}
\usepackage{color}
\usepackage{epsfig, epsf, graphicx}
\usepackage{animate}
\usepackage{pstricks, pst-node, psfrag}
\usepackage{bm}
\usepackage{amsmath}
\usepackage{amssymb}
\usepackage{verbatim,enumerate}
\usepackage{rotating, lscape}
\usepackage{diagbox}
\usepackage[doublespacing]{setspace}
\usepackage{tikz}
\usetikzlibrary{arrows,calc,tikzmark}
\usepackage[colorlinks=true,linkcolor=blue,citecolor=blue]{hyperref}%
\usepackage{float}
\usepackage[small]{caption}
\usepackage{subcaption}
\usepackage[square]{natbib}
\usepackage{lipsum}
\usepackage{threeparttable}
\usepackage{multirow}
\usepackage[toc,page]{appendix}
\def\diag{\hbox{diag}}

\def\boxit#1{\vbox{\hrule\hbox{\vrule\kern6pt
			\vbox{\kern6pt#1\kern6pt}\kern6pt\vrule}\hrule}}

\def\bse{\begin{eqnarray*}}
	\def\ese{\end{eqnarray*}}
\def\be{\begin{eqnarray}}
\def\ee{\end{eqnarray}}
\def\bq{\begin{equation}}
\def\eq{\end{equation}}
\def\bse{\begin{eqnarray*}}
	\def\ese{\end{eqnarray*}}

\begin{document}

\thispagestyle{empty} \baselineskip=28pt \vskip 5mm
\begin{center} {\Huge{\bf Sparse Functional Boxplots for Multivariate Curves}}
	
\end{center}

\baselineskip=12pt \vskip 10mm

\begin{center}\large
Zhuo Qu and Marc G.~Genton\footnote[1]{
\baselineskip=10pt Statistics Program,
King Abdullah University of Science and Technology,
Thuwal 23955-6900, Saudi Arabia.\\
E-mail: zhuo.qu@kaust.edu.sa, marc.genton@kaust.edu.sa\\
This research was supported by the
King Abdullah University of Science and Technology (KAUST).}
\end{center}

\baselineskip=17pt \vskip 10mm \centerline{\today} \vskip 15mm

\begin{center}
{\large{\bf Abstract}}
\end{center}
This paper introduces the sparse functional boxplot and the intensity sparse functional boxplot as practical exploratory tools. Besides being available for complete functional data, they can be used in sparse univariate and multivariate functional data. The sparse functional boxplot, based on the functional boxplot, displays sparseness proportions within the 50\% central region. The intensity sparse functional boxplot indicates the relative intensity of fitted sparse point patterns in the central region. The two-stage functional boxplot, which derives from the functional boxplot to detect outliers, is furthermore extended to its sparse form. We also contribute to sparse data fitting improvement and sparse multivariate functional data depth. In a simulation study, we evaluate the goodness of data fitting, several depth proposals for sparse multivariate functional data, and compare the results of outlier detection between the sparse functional boxplot and its two-stage version. The practical applications of the sparse functional boxplot and intensity sparse functional boxplot are illustrated with two public health datasets. Supplementary materials and codes are available for readers to apply our visualization tools and replicate the analysis.

\baselineskip=14pt

\par\vfill\noindent
{\bf Some key words:} Depth, Missing values, Multivariate functional data, Outlier detection, Sparse functional data, Visualization

\clearpage\pagebreak\newpage \pagenumbering{arabic}
\baselineskip=26pt

\section{Introduction}
\label{sec:intro}
Functional data analysis (\citeauthor{ramsay1991some} \citeyear{ramsay1991some}) regards each observation unit as a function of an index displayed as a curve or image. Descriptive statistics, such as median (\citeauthor{sun2012functional} \citeyear{sun2012functional}, \citeauthor{qu2021robust} \citeyear{qu2021robust}) and outliers (\citeauthor{dai2018directional} \citeyear{dai2018directional}), need to be determined before the functional data can be displayed for exploratory data analysis. Ordering the samples of curves and images directly is difficult.~Therefore, many functional depths have been proposed that establish the center of the functional data sample and then order them from the center outwards. Available functional depths in a marginal perspective are the integrated depth (\citeauthor{ibrahim2009missing} \citeyear{ibrahim2009missing}), random projection depth (\citeauthor{cuevas2007robust} \citeyear{cuevas2007robust}), random Tukey depth (\citeauthor{cuesta2008random} \citeyear{cuesta2008random}), band and modified band depth  (\citeauthor{lopez2009concept} \citeyear{lopez2009concept}), half-region depth and modified half-region depth (\citeauthor{lopez2011half} \citeyear{lopez2011half}), functional spatial depth (\citeauthor{sguera2014spatial} \citeyear{sguera2014spatial}), and extremal depth (\citeauthor{narisetty2016extremal} \citeyear{narisetty2016extremal}). Moreover, multivariate functional depths have been proposed based on two generalizations: from functional to multivariate functional scenarios and from multivariate to multivariate functional scenarios. Examples of the first idea include the weighted average of the marginal functional depths (\citeauthor{ieva2013depth} \citeyear{ieva2013depth}) and the simplicial and modified simplicial band depths (\citeauthor{lopez2014simplicial} \citeyear{lopez2014simplicial}). An example of the second idea is the multivariate functional halfspace depth (\citeauthor{claeskens2014multivariate} \citeyear{claeskens2014multivariate}). 

With the intensive development of different depth notions, various tools have been developed for visualizing functional data and detecting outliers.~\citeauthor{hyndman2010rainbow} (\citeyear{hyndman2010rainbow}) made use of the first two robust functional principal component scores and presented functional versions of the bagplot and the highest density region plot. With the modified band depth giving the ranks among data, \citeauthor{sun2011functional} (\citeyear{sun2011functional}) proposed the functional boxplot, a potent analog to the classical boxplot (\citeauthor{tukey1975mathematics} \citeyear{tukey1975mathematics}), for visualizing functional data. For functional data with a dependence structure, \citeauthor{sun2012adjusted} (\citeyear{sun2012adjusted}) provided an adaptive way of determining the outlier selection factor in the adjusted functional boxplot.~\citeauthor{arribas2014shape} (\citeyear{arribas2014shape}) explored the relationship between modified band depth and modified epigraph index (\citeauthor{lopez2011depth} \citeyear{lopez2011depth}) and proposed the outliergram to visualize and detect shape outliers among functional data. \citeauthor{genton2014surface} (\citeyear{genton2014surface}) introduced a surface boxplot based on the modified volume depth and developed an interactive surface boxplot tool for the visualization of samples of images. \citeauthor{mirzargar2014curve} (\citeyear{mirzargar2014curve}) extended the notion of depths from functional data to curves, presenting the curve boxplot for visualizing ensembles of 2D and 3D curves. \citeauthor{dai2018functional} (\citeyear{dai2018functional}) developed a two-stage functional boxplot for multivariate curves, combining directional outlyingness (\citeauthor{dai2018directional} \citeyear{dai2018directional}) and a classical functional boxplot in the outlier detection procedure. Additionally, \citeauthor{dai2018multivariate} (\citeyear{dai2018multivariate}) introduced the magnitude-shape (MS) plot for visualizing both the magnitude and shape outlyingness of multivariate functional data. \citeauthor{yao2020trajectory} (\citeyear{yao2020trajectory}) proposed the trajectory functional boxplot and the modified simplicial band depth (MSBD) versus wiggliness of directional outlyingness (WO) plot for visualizing trajectory functional data. For more visualization examples, see the recent review of \citeauthor{genton2020functional} (\citeyear{genton2020functional}).

The above visualization techniques can only be applied to samples of curves measured on fine and common grids. In practice, the grids are not always fine or common; rather, curves are sometimes observed on sparse or irregularly spaced time points. 
\begin{figure}[b!]
    \centering
    \includegraphics[width = 0.5\textwidth, height = 8cm]{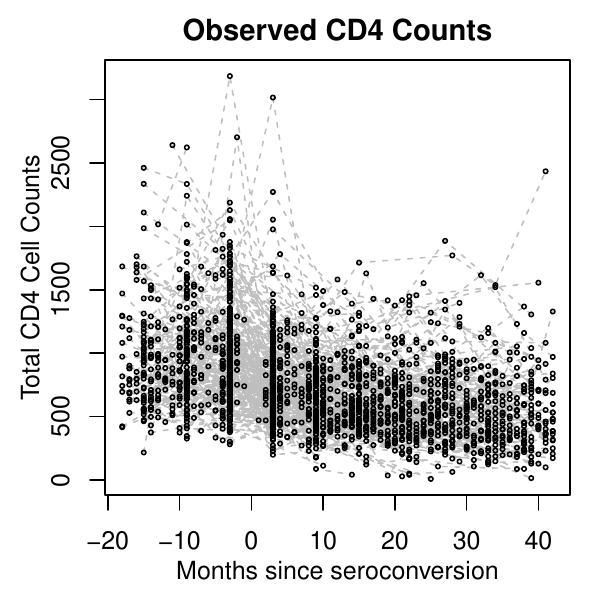}
    \caption{Observed CD4 cell counts from 366 patients (black solid points) measured from $18$ months before to $42$ months after seroconversion (the time period during which a specific antibody develops and becomes detectable in the blood).}
    \label{cd4}
\end{figure} 
Figure \ref{cd4} shows a longitudinal study of CD4 cell counts (\citeauthor{goldsmith2013corrected} \citeyear{goldsmith2013corrected}) per milliliter of blood to track the progress of HIV. Overall, there are $366$ patients, with between $1$ and $11$ observations per subject and a mean of $5$, yielding 1888 data points.
\begin{figure}[!b]
    \centering
    \includegraphics[width=0.92\textwidth, height = 7cm]{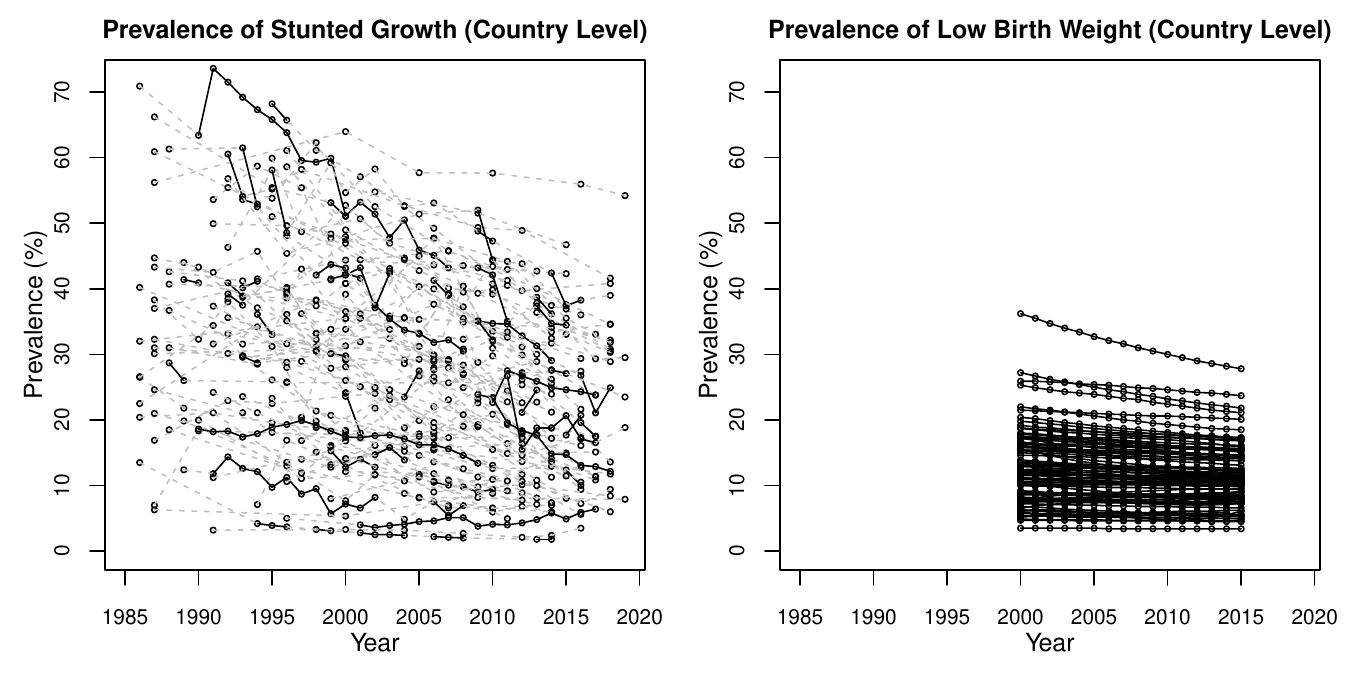}
    \caption{The observed prevalence of stunted growth and prevalence of low birth weight for 77 countries from 1985 to 2019. Observations are joined with solid black lines if observed continuously; otherwise, joined with gray dashed lines.}
   \label{org_mal}
\end{figure}
Figure \ref{org_mal} presents a dataset of malnutrition metrics, consisting of the prevalence of stunted growth and low birth weight, between 1985 to 2019 for 77 countries. Sparseness appears randomly for the first variable, stunted growth, and values for low birth weight are only available for years 2000 to 2015. Applying visualization tools to sparse functional data is difficult but necessary in order to visualize sparse multivariate functional data.

Various ideas have been proposed to deal with missingness in the notion of depth in the univariate functional case. A first approach is to fit the sparse functional data before applying current univariate depths and a second approach is to revise the notions of depth for the sparse functional data case. One example of the first approach is in \citeauthor{lopez2011depth} (\citeyear{lopez2011depth}), in which the ideas of univariate functional principal component analysis (UFPCA, \citeauthor{yao2005functional} \citeyear{yao2005functional}) are applied for sparse data fitting and the modified band depth is used to order the fitted data. Other available methods for fitting sparse functional data include UFPCA under different frameworks (\citeauthor{james2000principal} \citeyear{james2000principal}, \citeauthor{liu2017functional} \citeyear{liu2017functional}), B-spline models (\citeauthor{thompson2008bayesian} \citeyear{thompson2008bayesian}), and covariance estimation (\citeauthor{xiao2018fast} \citeyear{xiao2018fast}). It is worth noting that \citeauthor{goldsmith2013corrected} (\citeyear{goldsmith2013corrected}) proposed a bootstrap procedure to improve curve estimates by accounting for both uncertainty in UFPCA decompositions and model-based estimates based on UFPCA estimates in \citeauthor{yao2005functional} (\citeyear{yao2005functional}). Additionally, pointwise and simultaneous confidence bands that account for both model- and decomposition-based variability were constructed. One example of the second approach is in \citeauthor{sguera2021notion} (\citeyear{sguera2021notion}), where the idea of depth is extended by incorporating both the curve estimation and its confidence bands (\citeauthor{goldsmith2013corrected} \citeyear{goldsmith2013corrected}) into the depth analysis, which can be applied to any univariate functional depth. 

Similar methods have been generalized to fit sparse data in the multivariate functional setting. \citeauthor{zhou2008joint} (\citeyear{zhou2008joint}) modeled paired longitudinal data with principal components and fitted them by penalized splines under the mixed-effects model framework. However, their method is limited to only two variables when the data are either observed or missing simultaneously. \citeauthor{happ2018multivariate} (\citeyear{happ2018multivariate}) proposed a multivariate functional principal component analysis (MFPCA) defined on different time grids, which is also suitable for data with missing values. \citeauthor{li2020fast} (\citeyear{li2020fast}) derived a fast algorithm for fitting sparse multivariate functional data via estimating the multivariate covariance function with tensor product B-spline. However, a bootstrap procedure was not considered in the sparse multivariate functional data to improve curve estimates. Still, the depth of sparse multivariate functional data has not been considered in detail. 

The novel contributions in this paper are threefold. First, we implement a bootstrap procedure to improve the MFPCA fit (\citeauthor{happ2018multivariate} \citeyear{happ2018multivariate}). Second, we propose a revised depth in the sparse multivariate functional case to determine the order of sparse multivariate functional data. Third, we develop visualization tools for both marginal and joint sparse functional data. The sparse functional boxplot mainly displays the percentage of sparse points within the central region, whereas the intensity sparse functional boxplot highlights the intensity of the sparse points in the central region.
 
The remainder of the paper is organized as follows.~The procedure of fitting sparse multivariate functional data and computing depths for sparse multivariate functional data are considered in Section \ref{sec2}. Sparse and intensity sparse functional boxplots, together with their two-stage forms, are presented in Section \ref{sec3}. The depth selection and outlier detection performance for different visualization tools are demonstrated in Section \ref{sec4}. Applications to the aforementioned CD4 and malnutrition data are presented in Section \ref{sec5}. The paper ends with a discussion in Section \ref{sec6}.
\section{Ordering Sparse Multivariate Functional Data}
\label{sec2}
We consider a two-step procedure to order sparse multivariate functional data: first we generalize the bootstrap procedure (\citeauthor{goldsmith2013corrected} \citeyear{goldsmith2013corrected}) from UFPCA to MFPCA (\citeauthor{happ2018multivariate} \citeyear{happ2018multivariate}) to generate improved curve estimates and confidence bands; second, we consider various notions of depth for sparse multivariate functional data. The depth notions we consider are either from the direct generalization of the univariate revised functional depth (\citeauthor{sguera2021notion} \citeyear{sguera2021notion}) to multivariate functional data, or our new idea described in Section \ref{rd}.
\subsection{Fitting Sparse Multivariate Functional Data}
\label{s2}
\subsubsection{Data Structure and Notation}
This paper considers multivariate functional data. A multivariate functional random variable $\bm{Y}$ (\citeauthor{hsing2015theoretical} \citeyear{hsing2015theoretical}) is a random vector with values in an infinite-dimensional space. A well-received model of multivariate functional data is to consider paths of a stochastic process taking values in some Hilbert space, $\mathcal{H}$, of functions defined on some set $\mathcal{T}$. First we consider a $p$-variate ($p \in \mathbb{Z}_{+}$) stochastic process $\bm{Y}(\bm{t}) =({Y}^{(1)}(t^{(1)}),\ldots, {Y}^{(p)}(t^{(p)}))^\top$ with $\bm{t}^\top:=(t^{(1)},\ldots,t^{(p)}) \in \mathcal{T}:=\mathcal{T}_1 \times \cdots \times \mathcal{T}_p$. Note that $\bm{t}$ is a $p$-dimensional vector, with its element $t^{(j)}$ being a random time and independent of all other random variables. Each element $Y^{(j)}(t^{(j)})$ ($j = 1,\ldots, p$) is defined on the domain $\mathcal{T}_j$, where the $\mathcal{T}_j$s are compact sets in $\mathbb{R}$ with finite Lebesgue measure. Briefly speaking, $Y^{(j)}(t^{(j)})$: $\mathcal{T}_j \to \mathbb{R}$ is assumed to be square-integrable in $\mathcal{T}_j$, expressed as $L^2(\mathcal{T}_j)$. Second, we consider the $p$-dimensional functional data $\bm{Y}=\{\bm{Y}(\bm{t})\}_{\bm{t} \in \mathcal{T}}$ and we have $\bm{Y} \in \mathcal{H}$, where the space $\mathcal{H}:= L^2(\mathcal{T}_1)\times \cdots \times L^2(\mathcal{T}_p)$.

For $\bm{s}, \bm{t} \in \mathcal{T}$, define the matrix of covariances $\bm{C}(\bm{s}, \bm{t}):= {\rm cov}\{\bm{Y}(\bm{s}), \bm{Y}(\bm{t})\}$ with elements $C_{ij}(s^{(i)}, t^{(j)}):={\rm cov} \{Y^{(i)}(s^{(i)}), Y^{(j)}(t^{(j)})\}$ for $s^{(i)} \in \mathcal{T}_i$ and $t^{(j)} \in \mathcal{T}_j$. Define the covariance operator $\Gamma: \mathcal{H} \to \mathcal{H}$ with the $j$th element of $\Gamma f$ for $f\in \mathcal{H}$ given by $(\Gamma f)^{(j)}(t^{(j)}):=\sum_{i=1}^p \int_{\mathcal{T}_i}C_{ij}(s^{(i)}, t^{(j)})f^{(i)}(s^{(i)}){\rm d}s^{(i)}$ for $t_j \in \mathcal{T}_j$. By the Hilbert-Schmidt Theorem (\citeauthor{renardy2006introduction} \citeyear{renardy2006introduction} p. 253-262), it follows that there exists a complete orthonormal basis of eigenfunctions $\bm{\psi}_m = (\psi_m^{(1)},\ldots,\psi_m^{(p)})^\top \in \mathcal{H}$, $m \in \mathbb{N}$ of $\Gamma$ such that $\Gamma \bm{\psi}_m = \nu_m \bm{\psi}_m$ and $\lim\limits_{m \to +\infty}\nu_m = 0$. Here, $\nu_m$ is a sequence of non-zero real eigenvalues such that $\nu_1 \geq \nu_2 \geq \cdots \geq 0$. In Proposition 2 of \citeauthor{happ2018multivariate} (\citeyear{happ2018multivariate}), the covariance operator $\Gamma$ is a linear, self-adjoint and positive operator. Let $\sum_{m=1}^{\infty} \nu_{m}{\psi}_{m}^{(j)}(s^{(j)}){\psi}_{m}^{(j)}(t^{(j)})$ be the spectral decomposition of $C_{jj}(s^{(j)}, t^{(j)})$. Based on the property of $\Gamma$ and the decomposition of covariance elements, a multivariate Karhunen–Lo\`{e}ve representation (\citeauthor{happ2018multivariate} \citeyear{happ2018multivariate}) for $\bm{Y}(\bm{t})$ is $\bm{Y}(\bm{t}) = \bm{\mu} (\bm{t}) + \sum_{m=1}^{\infty} \rho_{m}\bm{\psi}_{m}(\bm{t})$ for $\bm{t} \in \mathcal{T}$, $\bm{\mu}(\bm{t}) := \textrm{E}\{\bm{Y}(\bm{t})\}$ is the mean function $\bm{\mu}$ evaluated at time $\bm{t}$, with the element $\mu^{(j)}(t^{(j)})= \textrm{E}\{{Y}^{(j)}(t^{(j)})\}$ for $j=1,\ldots,p$, and $\bm{\psi}_{m}(\bm{t})$ is the $m$th~($m \in \mathbb{N}$) eigenfunction evaluated at time $\bm{t}$. Here, $\rho_{m}=\sum_{j=1}^{p}\int_{\mathcal{T}_j}\{Y^{(j)}(t^{(j)}) - \mu^{(j)}(t^{(j)})\}\psi_m^{(j)}(t^{(j)}){\rm d}t^{(j)}$ are zero mean random variables with ${\rm cov}(\rho_m, \rho_n) = \nu_m$ if $m=n$, and ${\rm cov}(\rho_m, \rho_n) = 0$ if $m\neq n$ for $m,n \in \mathbb{N}$.

In the following, let $\bm{Y}_1,\ldots, \bm{Y}_N$ be a set of independent observations of $\bm{Y}$.~In practice, we observe the functions $\bm{Y}_i(\bm{t})$~($i=1,\ldots,N$) with error $\bm{\epsilon}_i(\bm{t}) = (\epsilon_i^{(1)}(t^{(1)}), \ldots, \epsilon_i^{(p)}(t^{(p)}))^\top$, and the element $\epsilon_i^{(j)}(t^{(j)})^\top\stackrel{i.i.d.}{\sim} \mathcal{N}(0, \sigma_j^2)$. Moreover, the functions $\bm{Y}_i(\bm{t}_i)$ are observed on sparse finite grids at the subject and element level, that is, the $j$th ($j=1,\ldots,p$) element $t_i^{(j)}$ of $\bm{t}_i~ (\bm{t}_i \in \mathcal{T})$ can vary per curve. Let the observed functions with measurement errors and the sparseness be $\widetilde{\bm{Y}}_i(\bm{t}_i)$ such that $\widetilde{\bm{Y}}_i(\bm{t}_i) = \bm{Y}_i(\bm{t}_i) + \bm{\epsilon}_i(\bm{t}_i)$.

Then, the estimation of observed functions $\widetilde{\bm{Y}}_i(\bm{t}_i)$ can be approximated by the multivariate truncated Karhunen-Lo\`{e}ve expansions (\citeauthor{happ2018multivariate} \citeyear{happ2018multivariate}) $\overline{\bm{Y}}_i(\bm{t}_i)$ such that $\overline{\bm{Y}}_i(\bm{t}_i) = \bm{\mu} (\bm{t}_i) + \sum_{m=1}^{M} \rho_{i, m}\bm{\psi}_{m}(\bm{t}_i) + \bm{\epsilon}_i(\bm{t}_i)$. Here, $M$ is the minimal number of eigenfunctions that explain 99\% variability in the observed curves. Define $\bm{\Psi} = (\bm{\psi}_1, \ldots, \bm{\psi}_M)$ the collection of the first $M$ eigenfunctions, $\bm{V} = \diag\{\nu_1,\ldots, \nu_M\}$, $\bm{\Sigma} = \diag\{\sigma^2_1,\ldots, \sigma^2_p\}$. We name the decomposition objects $\bm{\theta} = \{M, \bm{\mu}, \bm{\Psi}, \bm{V}, \bm{\Sigma}\}$. Let the eigenscores for function $\bm{Y}_i$ be $\bm{\rho}_i=\{\rho_{i,m}: m = 1, \ldots, M\}$, the estimation of $\bm{\rho}_{i}$ is regarded as an expectation based on the objects estimated in the MFPCA decomposition, $\widehat{\bm{\theta}}$. Then, the estimation of $\overline{\bm{Y}}_i(\bm{t}_i)$ is seen as the expectation of $\widetilde{\bm{Y}}_i(\bm{t}_i)$ based on the estimated $\widehat{\bm{\theta}}$ and $\widehat{\bm{\rho}}_i$. Take a random $\bm{t} = (t^{(1)}, \ldots, t^{(p)})^\top \in \mathcal{T}$. The estimated $j$th component of the $i$th observation at $t^{(j)}$ is expressed as below:
\begin{equation}
\begin{footnotesize}
\widehat{{Y}}^{(j)}_{\widehat{\bm{\theta}}, i}(t^{(j)}) = {\textrm{E}}\Big\{\widetilde{Y}^{(j)}_i(t^{(j)})| \widehat{\bm{\rho}}_{\widehat{\bm{\theta}},i}, \widehat{\bm{\theta}}\Big\} = \widehat{{\mu}}^{(j)} (t^{(j)}) + \sum_{m=1}^{M} \widehat{\rho}_{i, m}\widehat{{\psi}}^{(j)}_{m}(t^{(j)}),~i = 1,\ldots,N, ~j = 1, \ldots, p, ~t^{(j)} \in \mathcal{T}^{(j)}.
\end{footnotesize}
\label{model-based-estimate}
\end{equation}
The MFPCA estimation is available in \textit{MFPCA} package (\citeauthor{happ2018mfpca} \citeyear{happ2018mfpca}) in R (\citeauthor{ihaka1996r} \citeyear{ihaka1996r}).

\subsubsection{Bootstrap Improved Estimation and Confidence Bands}
The MFPCA fit is a model-based estimate without taking into account the uncertainty in the decomposition objects. Hence, we implement the bootstrap (\citeauthor{goldsmith2013corrected} \citeyear{goldsmith2013corrected}) to combine the model-based conditional estimates across the distribution of decompositions. 

Assume we implement the bootstrap with $B = 100$ replicates. We randomly select $N$ functions from observations $\bm{Y}_1,\ldots, \bm{Y}_N$ with replacement, and obtain the model-based estimate in (\ref{model-based-estimate}). The bootstrap MFPCA (BMFPCA) fit is obtained from the iterated expectation and we provide the estimated $j$th component of the $i$th observation at $t^{(j)}$ as below:
\begin{equation}
    \widehat{{Y}}^{(j)}_i(t^{(j)})=\mathbb {E}_{\widehat{\bm{\theta}}}\Big[\textrm{E}_{\widetilde{\bm{Y}}^{(j)}_i|\widehat{\bm{\theta}}}\Big\{\widetilde{{Y}}^{(j)}_i(t^{(j)})| \widehat{\bm{\rho}}_{\widehat{\bm{\theta}},i}, \widehat{\bm{\theta}}\Big\} \Big],~~~i=1,\ldots,N, ~j = 1,\ldots, p,~t^{(j)} \in \mathcal{T}^{(j)}.
    \label{eq}
\end{equation}

    Like \citeauthor{happ2018multivariate} (\citeyear{happ2018multivariate}), we do not assume any distribution for $\rho_{i, m}$. Thus, we use the naive bootstrap to obtain the confidence bands. Given the confidence level $\alpha$, for each fixed time grid and the component of the curve, we take the $(1-\frac{\alpha}{2})$th percentile across the bootstrap as the confidence upper bound, and the $\frac{\alpha}{2}$th percentile across the bootstrap as the confidence lower bound. Here, we define the $j$th component of the $i$th upper bound at $t^{(j)}$ as ${Y}^{(j)}_{ub,i}(t^{(j)}): = q_{1 - \frac{\alpha}{2}}(\widehat{Y}^{(j)}_{\widehat{\bm{\theta}}_1,i}(t^{(j)}), \ldots, \widehat{Y}^{(j)}_{\widehat{\bm{\theta}}_B,i}(t^{(j)}))$, where $q_{\beta}(A)$ denotes the $\beta$th percentile of the set $A$. Similarly, the $j$th component of the $i$th lower bound at $t^{(j)}$ is ${Y}^{(j)}_{lb,i}(t^{(j)}):= q_{\frac{\alpha}{2}}(\widehat{Y}^{(j)}_{\widehat{\bm{\theta}}_1,i}(t^{(j)}), \ldots, \widehat{Y}^{(j)}_{\widehat{\bm{\theta}}_B,i}(t^{(j)}))$. We can easily obtain $\bm{Y}_{ub,i}(\bm{t})=({Y}^{(1)}_{ub,i}(t^{(1)}),\ldots,{Y}^{(p)}_{ub,i}(t^{(p)}))^\top$ and $\bm{Y}_{lb,i}(\bm{t})=({Y}^{(1)}_{lb,i}(t^{(1)}),\ldots,{Y}^{(p)}_{lb,i}(t^{(p)}))^\top$ with each component given, and the confidence bands bounded by $\bm{Y}_{ub,i}$ and $\bm{Y}_{lb,i}$.

\subsection{Notions of Depth for Sparse Multivariate Functional Data}
\label{s3}
We take multivariate functional halfspace depth (MFHD, \citeauthor{hubert2015multivariate} \citeyear{hubert2015multivariate}) as an example to represent depth revision in sparse settings in later sections; in principle, any reasonable multivariate functional depth can be used. Since multivariate functional depth (MFD) requires common time grids for all variables, without loss of generality, we let $\mathcal{T}_j = \widetilde{\mathcal{T}}$ for $j=1,\ldots,p$, where $\widetilde{\mathcal{T}}$ is a compact set in $\mathbb{R}$. Later, the revised depth can be proposed by applying MFHD to the fitted data (see \citeauthor{lopez2011depth} \citeyear{lopez2011depth} in the univariate sparse functional case), or by taking the depths of fitted data and their confidence bands into account. There are several implementations in the latter case, see section \ref{rd} for details. 

\subsubsection{Conventional Depth for Multivariate Functional Data}
For $\bm{\mathcal{X}}$ a random vector on $\mathbb{R}^p$, let $D(\cdot;F_{\bm{\mathcal{X}}}): \mathbb{R}^p \xrightarrow{}[0,1]$ be a statistical depth function (\citeauthor{zuo2000general} \citeyear{zuo2000general}) for the probability distribution of $\bm{\mathcal{X}}$ with associated cumulative distribution function (cdf) $F_{\bm{\mathcal{X}}}$. Accordingly, the depth region $D_{\beta}(F_{\bm{\mathcal{X}}})$ at level $\beta\geq 0$, is defined as $D_{\beta}(F_{\bm{\mathcal{X}}})=\{\bm{X}\in \mathbb{R}^p: D(\bm{X}; F_{\bm{\mathcal{X}}})\geq \beta \}$. We let $C^p(\widetilde{\mathcal{T}})$ be continuous paths for the $p$-variate continuous functions with $t \in \widetilde{\mathcal{T}}$. The definition of MFD combines the local depths at each time point and includes a weight function that may change with time. 

Consider such a $p$-variate stochastic process $\{\bm{\mathcal{X}}(t), t \in \widetilde{\mathcal{T}}\}$ on $\mathbb{R}^p$ with cdf $F_{\bm{\mathcal{X}}(t)}$ at each time point $t$ that generates continuous paths in $C^p(\widetilde{\mathcal{T}})$. 
Take an arbitary $\bm{Y} \in C^p(\widetilde{\mathcal{T}})$. Let $w$ be a weight function that integrates to one on the domain $\widetilde{\mathcal{T}}$. The MFD of $\bm{Y}$ is defined as ${MFD}(\bm{Y}; F_{\bm{\mathcal{X}}})=\int_{\widetilde{\mathcal{T}}} {D}(\bm{Y}(t);F_{\bm{\mathcal{X}}(t)})w(t)\textrm{d}t$. Here, if the weight is a constant, then $w(t) := w$; if $w(t)$ changes with time $t$, then $w(t): = {\rm vol}\{{ D}_\beta(F_{\bm{\mathcal{X}}(t)})\}/\int_{\widetilde{\mathcal{T}}} {\rm vol}\{{D}_\beta (F_{\bm{\mathcal{X}}(u)})\}\textrm{d}u$ is proportional to the volume of the depth region at time point $t$. The population halfspace depth (HD, \citeauthor{tukey1975mathematics} \citeyear{tukey1975mathematics}) at $\bm{Y}(t) \in \mathbb{R}^p$ is ${HD}(\bm{Y}(t);F_{\bm{\mathcal{X}}(t)})=\inf\limits_{\bm{u} \in \mathbb{R}^p,\|\bm{u}\|=1}P\{\bm{u}^\top {\bm{\mathcal{X}}}(t)\geq \bm{u}^\top \bm{Y}(t)\}.$ Thus, the population definition of MFHD is expressed as:
\begin{equation}
{MFHD}(\bm{Y}; F_{\bm{\mathcal{X}}})=\int_{\widetilde{\mathcal{T}}} { HD}(\bm{Y}(t);F_{\bm{\mathcal{X}}(t)})w(t)\textrm{d}t.
\end{equation}

In practice, instead of observing curves, one observes curve evaluations at a set of time points $t_1 < \cdots < t_L$ in $\widetilde{\mathcal{T}}=[t_1, t_L]$, not necessarily equidistant.~Consider a sample of multivariate functional observations ${\bm{\mathcal{X}}}^f(t_l)=\{{\bm{\mathcal{X}}}_1(t_l),\ldots,{\bm{\mathcal{X}}}_N(t_l); t_l \in \widetilde{\mathcal{T}}\}$ with at each time point $t$ cdf $F_{{\bm{\mathcal{X}}}(t),N}$.~The sample multivariate functional depth at $\bm{Y}\in C^p(\widetilde{\mathcal{T}})$ is defined with $t_0=t_1$, $t_{L+1}=t_{L}$ and $W_l=\int_{(t_{l-1}+t_l)/2}^{(t_{l}+t_{l+1})/2}\widetilde{w}(t)\textrm{d}t$, by
	${MFD}(\bm{Y};{\bm{\mathcal{X}}}^f)=\sum_{l=1}^{L} {D}(\bm{Y}(t_l);F_{{\bm{\mathcal{X}}}(t_l), N})W_l$. Similarly, $W_l = \widetilde{w}\cdot (t_{l+1}-t_{l-1})/2$ if $\widetilde{w}(t)$ is a constant, otherwise $W_l = {\rm vol} \{{ D}_\beta(F_{{\bm{\mathcal{X}}}(t_l),N})\}(t_{l+1}-t_{l-1})/[\sum_{l=1}^{L} {\rm vol}\{{ D}_\beta (F_{{\bm{\mathcal{X}}}(t_l),N})\}(t_{l+1}-t_{l-1})]$. The finite-sample halfspace depth (HD) is: ~\\${HD}(\bm{Y}(t_l); F_{{\bm{\mathcal{X}}}(t_l), N})=\frac{1}{N}\min\limits_{\bm{u}\in \mathbb{R}^p, \|\bm{u}\|=1}\#\{{\bm{\mathcal{X}}}_i(t_l), i=1,\ldots,N: \bm{u}^\top \bm{\mathcal{X}}_i(t_l)\geq \bm{u}^\top \bm{Y}(t_l)\}$, where $\#$ represents the number of counts.
Therefore, the finite sample definition of MFHD is:
\begin{equation}{MFHD}(\bm{Y};{\bm{\mathcal{X}}}^f)=\sum_{l=1}^{L} {HD}(\bm{Y}(t_l);F_{{\bm{\mathcal{X}}}(t_l), N})W_l. 
\end{equation}
\subsubsection{Revised Depth for Sparse Multivariate Functional Data}
\label{rd}
 We let $\widehat{\bm{Y}}^f=\{\widehat{\bm{Y}}_1,\ldots,\widehat{\bm{Y}}_N\}$ be a set including $N$ samples of fitted multivariate functional data over dense time grids. We also obtain a set of confidence upper bounds $\widehat{\bm{Y}}^f_{ub}=\{\widehat{\bm{Y}}_{ub,1},\ldots,\widehat{\bm{Y}}_{ub, N}\}$ and confidence lower bounds $\widehat{\bm{Y}}^f_{lb}=\{\widehat{\bm{Y}}_{lb,1},\ldots,\widehat{\bm{Y}}_{lb, N}\}$. We list the revised depth notion in the finite-sample version, and the population version can be analogously derived. 

Besides applying the current multivariate functional depths to the fitted data, we consider two additional ideas for revising the notion of depth for sparse multivariate functional data. They define the revised depth of $\bm{Y}_i$~($i=1,\ldots,N$) as the weighted average MFHD of the fitted data $\widehat{\bm{Y}}_i$, the confidence upper bound $\widehat{\bm{Y}}_{ub,i}$, and the confidence lower bound $\widehat{\bm{Y}}_{lb,i}$ in $\widehat{\bm{Y}}_{upd}: = \{\widehat{\bm{Y}}^f, \widehat{\bm{Y}}^f_{ub}, \widehat{\bm{Y}}^f_{lb}\}$ with different weights. Specifically, the first depth assigns $1/3$ to the fitted data, and to each of its confidence upper and lower bounds, which is the same as that in \citeauthor{sguera2021notion} (\citeyear{sguera2021notion}) in the univariate sparse functional case, while the second depth assigns $1/2$ to the fitted data and $1/4$ to its confidence upper and lower bounds, respectively. The weights in the second depth are new. 
We name the aforementioned revised depths the average-weight revised depth $RMFHD_{aw}$, and non-average-weight revised depth $RMFHD_{naw}$:
\begin{equation}
\footnotesize{
{RMFHD}_{type}(\widehat{\bm{Y}};\widehat{\bm{Y}}_{upd})=
\begin{cases}
    \frac{1}{3}{MFHD}(\widehat{\bm{Y}};\widehat{\bm{Y}}_{upd})+\frac{1}{3}{MFHD}(\widehat{\bm{Y}}_{ub};\widehat{\bm{Y}}_{upd})+\frac{1}{3}{MFHD}(\widehat{\bm{Y}}_{lb};\widehat{\bm{Y}}_{upd}), & \textrm{if~}{type = aw,} \\
    \frac{1}{2}{MFHD}(\widehat{\bm{Y}};\widehat{\bm{Y}}_{upd})+\frac{1}{4}{MFHD}(\widehat{\bm{Y}}_{ub};\widehat{\bm{Y}}_{upd})+\frac{1}{4}{MFHD}(\widehat{\bm{Y}}_{lb};\widehat{\bm{Y}}_{upd}), & \textrm{if~}{type = naw}. \\
\end{cases}
}
\label{md1}    
\end{equation}
We apply the conventional depth MFHD and its revised forms RMFHD to various data settings in simulations in Section \ref{datasetting}, and we identify the best depth according to the Spearman rank coefficient in Section \ref{simulation1}.

\section{Construction of Sparse Functional Boxplots}
\label{sec3}
The revised depths in Section \ref{sec2} make the visualization of sparse multivariate functional data possible.~The novel visualization tools coined sparse functional boxplot and intensity sparse functional boxplot not only keep the features of the original functional boxplot (\citeauthor{sun2011functional} \citeyear{sun2011functional}) but also display sparseness features. We further add a subsection illustrating the process of obtaining the two-stage (\citeauthor{dai2018functional} \citeyear{dai2018functional}) version of those two visualization tools. Without loss of generality, we use $\bm{Z}_i$ $(i=1,\ldots,N)$ to express the fitted data $\widehat{\bm{Y}}_i$ and the corresponding set $\bm{Z}^f:=(\bm{Z}_1, \ldots, \bm{Z}_N)$ in Section \ref{sec3} and the directional outlyingness is applied to only $\bm{Z}_i$ but not its confidence bands.

\subsection{Sparse Functional Boxplot}
The functional boxplot is constructed given the functional depth values for functional observations. Similar to the classical boxplot (\citeauthor{tukey1977exploratory} \citeyear{tukey1977exploratory}), it mainly displays five characteristics: the central region, the median, the outliers, the non-outlying maximal bound, and the non-outlying minimal bound. The $50\%$ central region is delimited by the envelope of the $50\%$ deepest curves from the sample set $\bm{Z}^f$.~In particular, the sample 50\% central region $C_{0.5}$ for the $j$th ($j = 1,\ldots, p$) component is
$C^{(j)}_{0.5} = \{(t, Z^{(j)}(t)): \min\limits_{r = 1, \ldots, \lceil N/2 \rceil} Z^{(j)}_{[r]}(t) \leq Z^{(j)}(t) \leq \max\limits_{r = 1, \ldots, \lceil N/2 \rceil} Z^{(j)}_{[r]}(t),~t \in \widetilde{\mathcal{T}}\},$ where $\lceil N/2 \rceil$ is the smallest integer not less than $N/2$, and $Z^{(j)}_{[r]}(t)$ is the $j$th component of the $r$th deepest curve evaluated at $t$ for $t \in \widetilde{\mathcal{T}}$. The upper bound of the central region evaluated at time $t$ is $Z^{(j)}_{ub, 0.5}(t) = \max\limits_{r = 1, \ldots, \lceil N/2 \rceil} Z^{(j)}_{[r]}(t)$, and the lower bound evaluated at $t$ is $Z^{(j)}_{lb, 0.5}(t) = \min\limits_{r = 1, \ldots, \lceil N/2 \rceil} Z^{(j)}_{[r]}(t)$.~We also define the range of $C^{(j)}_{0.5}$ at $t$ as the difference between the upper bound and lower bound that $R^{(j)}_{0.5}(t) := \max\limits_{r = 1, \ldots, \lceil N/2 \rceil} Z^{(j)}_{[r]}(t) - \min\limits_{r = 1, \ldots, \lceil N/2 \rceil} Z^{(j)}_{[r]}(t)$. The median $Z^{(j)}_{\lceil 1 \rceil}(t)$ is the curve with the highest depth.

Then, a curve $\bm{Z}_o$ is detected as an outlier, if the measurement of the $j$th component of $\bm{Z}_o$ is higher than the summation of 1.5 times the range of $C^{(j)}_{0.5}$ and the upper bound of $C^{(j)}_{0.5}(t)$ (or lower than the difference between the lower bound of $C^{(j)}_{0.5}$ and 1.5 times the range of $C^{(j)}_{0.5}(t)$) at some time $t$. That is $Z^{(j)}_o(t) > Z^{(j)}_{ub, 0.5}(t) + 1.5 R^{(j)}_{0.5}(t)$ or $Z^{(j)}_o(t) < Z^{(j)}_{lb, 0.5}(t) - 1.5 R^{(j)}_{0.5}(t)$. Let $S_o$ be the set of functional outliers. The non-outlying maximal (minimal) bound is established by connecting the maximal (minimal) points at all time indexes excluding the outliers. That is, $Z^{(j)}_{ub}(t) = \max\limits_{\bm{Z} \in \bm{Z}^f \setminus S_o} Z^{(j)}(t)$, and $Z^{(j)}_{lb}(t) = \min\limits_{\bm{Z} \in \bm{Z}^f \setminus S_o} Z^{(j)}(t)$ for $t\in \widetilde{\mathcal{T}}$.

The sparse functional boxplot, apart from displaying the aforementioned features, underlines the sparseness features in the median $Z^{(j)}_{\lceil 1 \rceil}(t)$, the 50\% central region $C^{(j)}_{0.5}$, and the detected outliers $S_o$. The median is drawn in black for the observed values and gray for the missing ones (Figure \ref{example}, 2nd row). At each time point $t \in \widetilde{\mathcal{T}}$ within $C^{(j)}_{0.5}$, we count the number of missing points $n^{(j)}_{ms}(t)$ and that of observed points $n^{(j)}_{obs}(t)$, and obtain the sparseness proportion $p^{(j)}_s(t)$, which is the number of sparse points divided by the total number of points $n^{(j)}_{ms}(t)/\{n^{(j)}_{ms}(t) + n^{(j)}_{obs}(t)\}$. Then, we define the proportion line such that: $l^{(j)}(t, p^{(j)}_s(t)) := Z^{(j)}_{ub, 0.5}(t) - p^{(j)}_s(t) R^{(j)}_{0.5}(t)$ for $t \in \widetilde{\mathcal{T}}$. It is derived from subtracting the range of the central region $R^{(j)}_{0.5}(t)$ times the sparseness proportion $p^{(j)}_s(t)$ from the upper bound of $C^{(j)}_{0.5}(t)$ at each time index $t$. Then, we display this percentage with a smoothed $l^{(j)}(t, p^{(j)}_s(t))$ over $\widetilde{\mathcal{T}}$ showing the observed proportion below and the sparseness proportion above. We fill the observed proportion area in magenta and the sparseness proportion area in gray. In addition, we plot a thin dotted cyan reference line, $l^{(j)}(t, 0.5)$, to represent the case of $p^{(j)}_s(t) = 0.5$ for all $t\in \widetilde{\mathcal{T}}$. For the detected outliers, the fitted missing values are represented by dashed gray, whereas the observed values are marked with dashed red.
\subsection{Intensity Sparse Functional Boxplot}
The sparse functional boxplot provides information about the sparse point proportion at each time, but we do not know the distribution of sparse points within the central region. To give a pattern of the intensity of sparse points, we construct the intensity sparse functional boxplot. 

In addition to the aforementioned characteristics of the functional boxplot, we display the intensity of fitted missing point patterns within the 50\% central region $C^{(j)}_{0.5}$, for any $j$th ($j = 1,\ldots, p$) component. Assuming that we have altogether $S$ original missing point data within $C^{(j)}_{0.5}$, we regard the fitted sparse points within $C^{(j)}_{0.5}$ as a spatial point pattern $\bm{u}^{(j)}_s:=\{(t_s, Z^{(j)}_s) \in C^{(j)}_{0.5}, s= 1, \ldots, S\}$ with $t_s$ the time, and $Z^{(j)}_s$ the fitted value inside the central region. Then we compute a kernel smoothed intensity function (\citeauthor{diggle1985kernel} \citeyear{diggle1985kernel}) from the point pattern.~The intensity estimate is corrected by dividing it by the convolution of the Gaussian kernel with the observation window. Thus, the sparseness intensity at a new point $\bm{u}^{(j)} \in C^{(j)}_{0.5}$ is 
\begin{equation}
    \lambda(\bm{u}^{(j)}) = e(\bm{u}^{(j)})\sum_{s=1}^{S} w_s \mathcal{K}(\bm{u}^{(j)}_s - \bm{u}^{(j)}),
\end{equation}
where $\mathcal{K}$ is the Gaussian smoothing kernel, $e(\bm{u}^{(j)})$ is an edge correction factor, and $w_s$ are the weights. The computation of such intensity is available in the R package \textit{spatstat.core} (\citeauthor{intensity} \citeyear{intensity}). By default, the intensity values of fitted sparse point patterns are expressed in estimated missing points per unit area.
\begin{figure}[!b]
    \begin{minipage}{0.33\textwidth}
     \centering
     \includegraphics[width = .95\linewidth, height = 5cm]{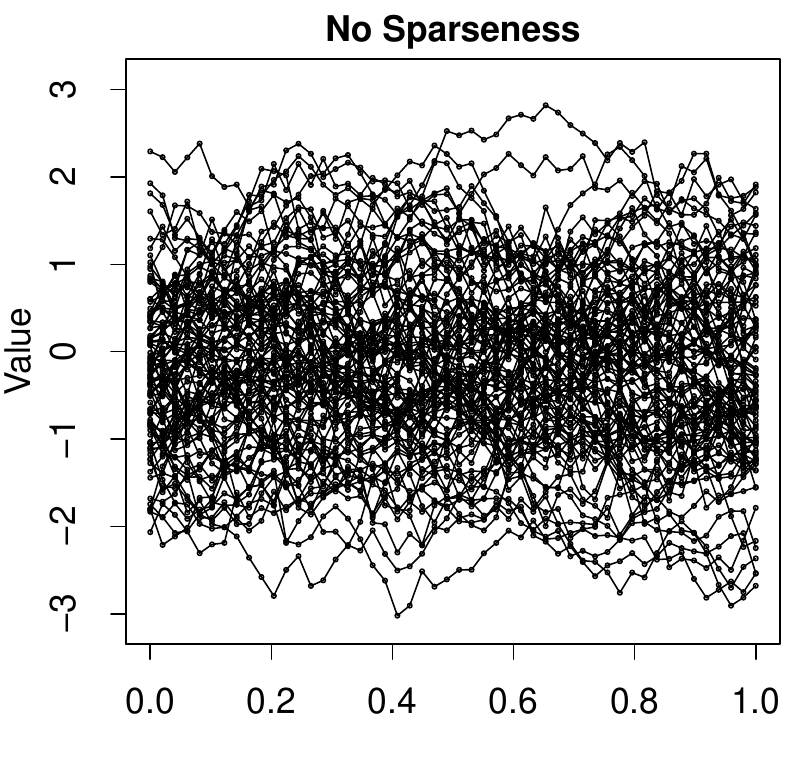}
   \end{minipage}\hfill
   \begin{minipage}{0.33\textwidth}
     \centering
     \includegraphics[width = 1\linewidth, height = 5cm]{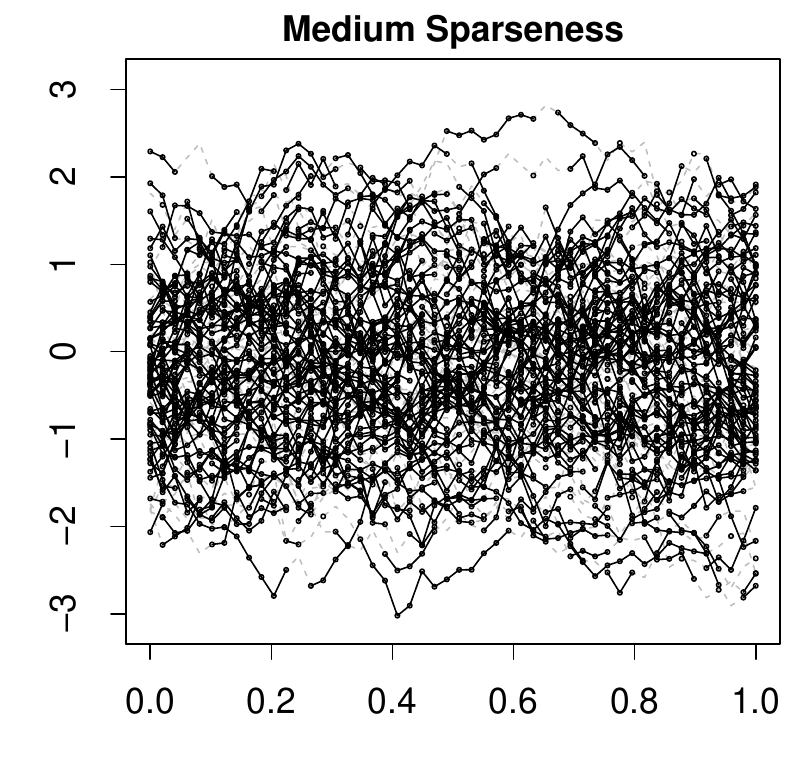}
   \end{minipage}
   \begin{minipage}{0.33\textwidth}
     \centering
     \includegraphics[width = 1\linewidth, height = 5cm]{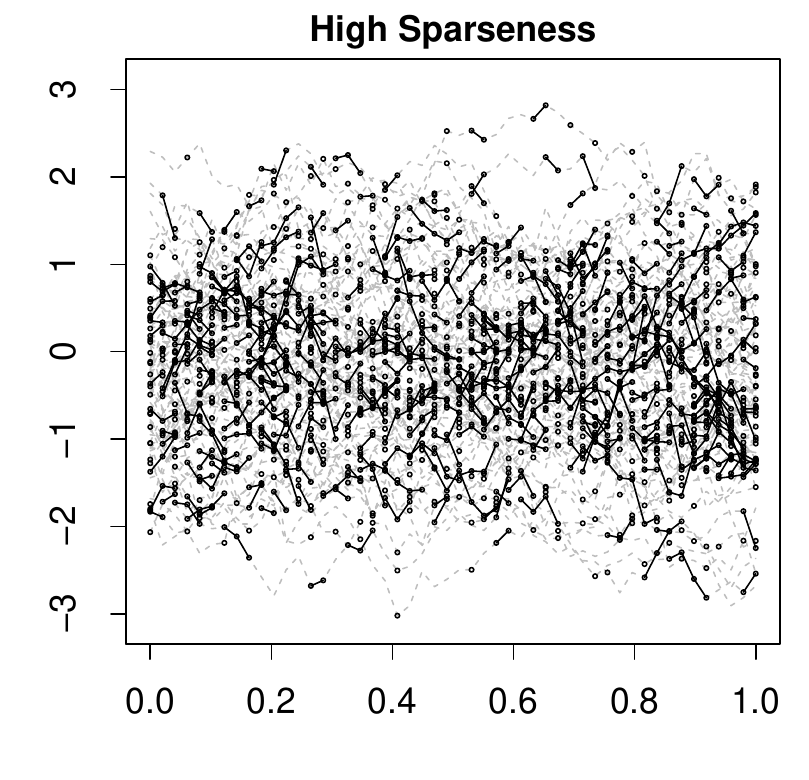}
   \end{minipage}
   \begin{minipage}{0.33\textwidth}
     \centering
     \includegraphics[width = .95\linewidth, height = 5cm]{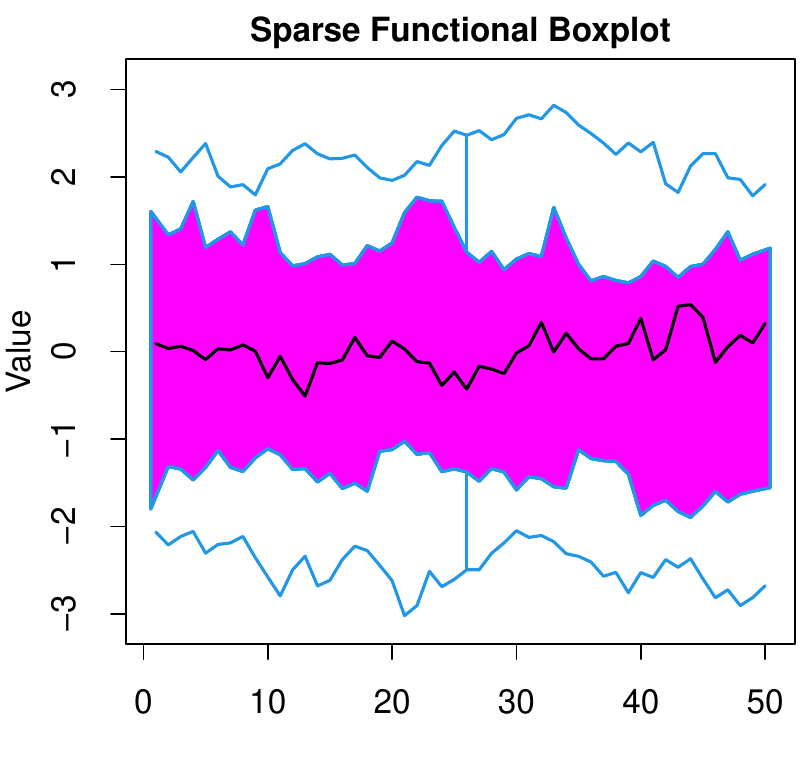}
   \end{minipage}\hfill
   \begin{minipage}{0.33\textwidth}
     \centering
     \includegraphics[width = 1\linewidth, height = 5cm]{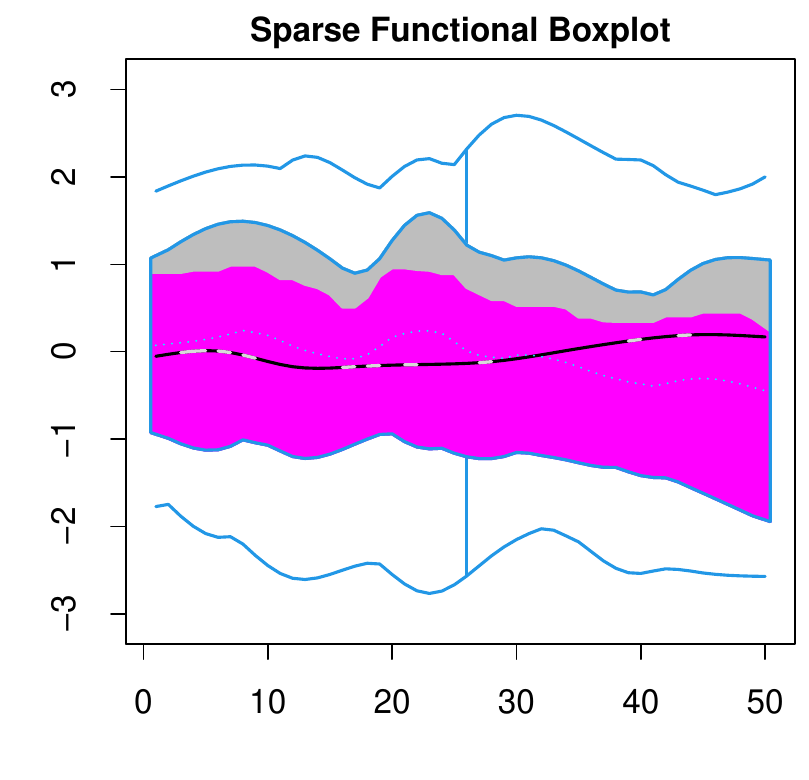}
   \end{minipage}
   \begin{minipage}{0.33\textwidth}
     \centering
     \includegraphics[width = 1\linewidth, height = 5cm]{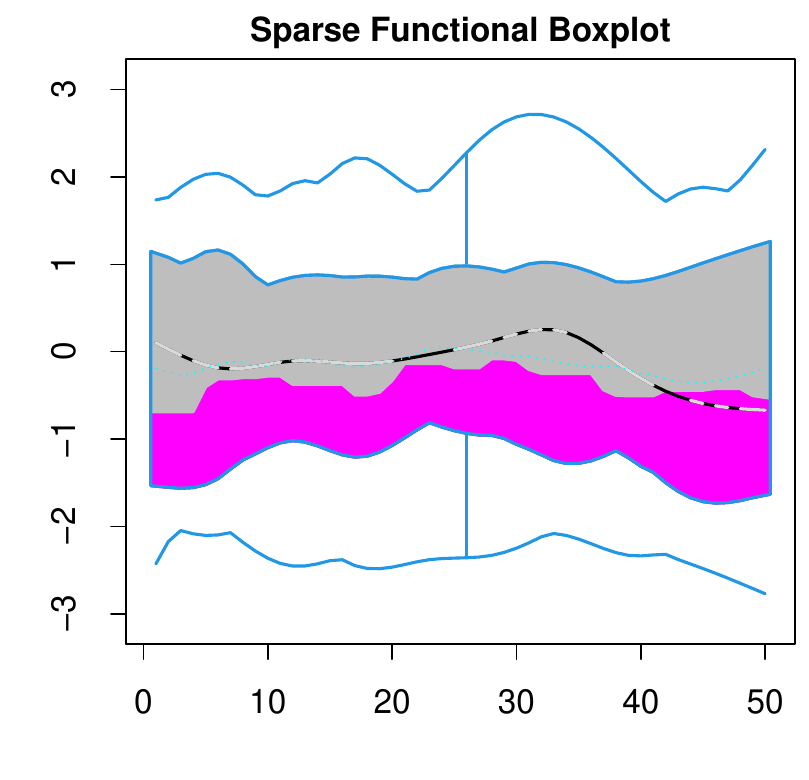}
   \end{minipage}
   
   \begin{minipage}{0.33\textwidth}
     \centering
     \includegraphics[width = .95\linewidth, height = 5cm]{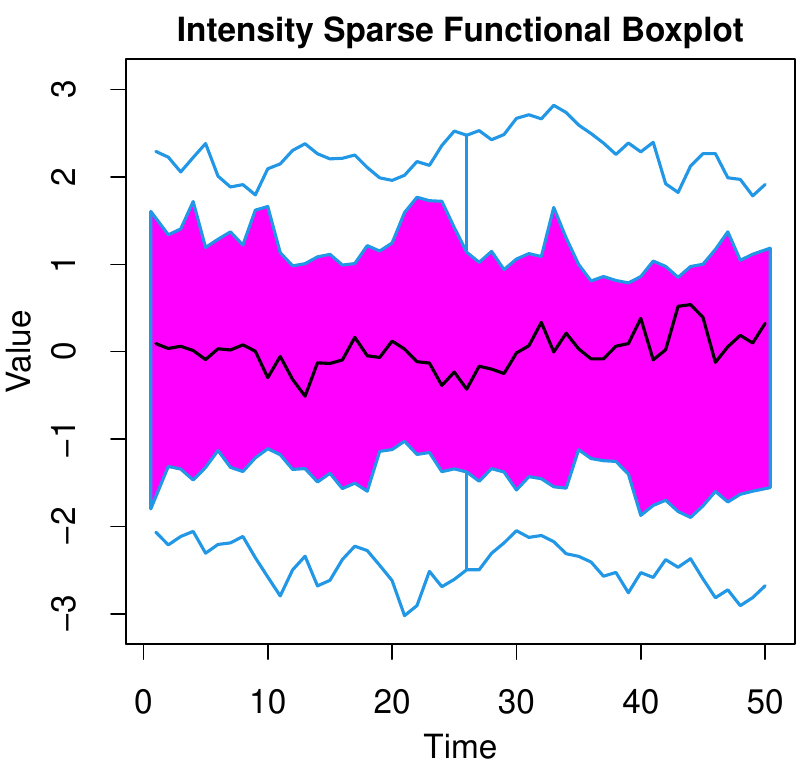}
   \end{minipage}\hfill
   \begin{minipage}{0.33\textwidth}
     \centering
     \includegraphics[width = 1\linewidth, height = 5cm]{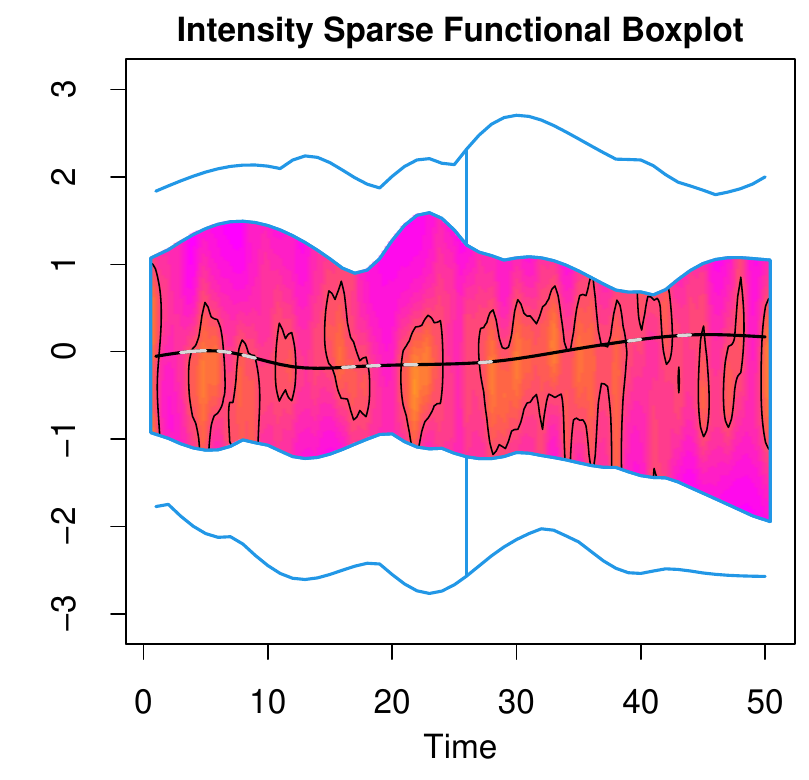}
   \end{minipage}
   \begin{minipage}{0.33\textwidth}
     \centering
     \includegraphics[width = 1.08\linewidth, height=5cm]{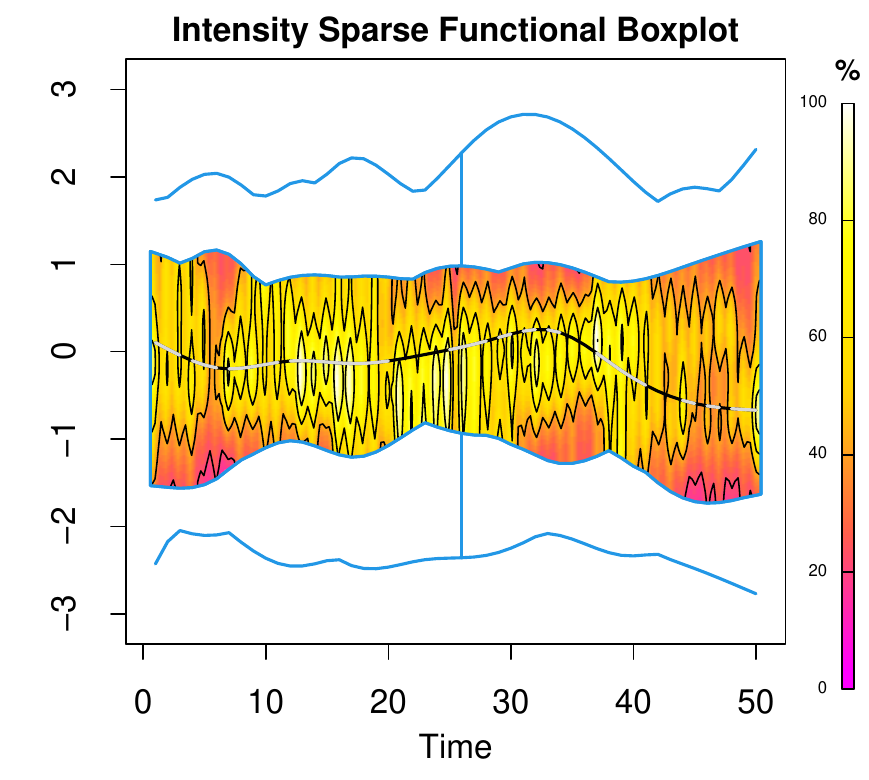}
   \end{minipage}
   \caption{The 1st row displays the univariate functional data from a Gaussian process with mean zero and exponential covariance function $C(t,s) = \exp\{-|t-s|\}$, with the corresponding sparse functional boxplots on the 2nd-row and the corresponding intensity sparse functional boxplots on the 3rd-row. Plots from left to right display three sparseness cases: no sparseness, medium sparseness (on average 20\% values are missing in each curve), and high sparseness (on average 60\% values are missing in each curve).}
   \label{example}
\end{figure}

Then, we normalize $\lambda(\bm{u}^{(j)})$ to $\lambda_{norm}(\bm{u}^{(j)}) := \frac{\lambda(\bm{u}^{(j)})}{\max\limits_{\bm{u}^{(j)}_s}\{\lambda(\bm{u}^{(j)}_s)\}}$.~To visualize the change of the relative intensity, we use a color scale between magenta for the case of zero intensity within the central region ($\lambda_{norm}(\bm{u}^{(j)}) = 0$), and white for the case of maximal intensity within the central region ($\lambda_{norm}(\bm{u}^{(j)}) = 1$). Furthermore, we can divide the intensity $\lambda(\bm{u}^{(j)})$ by the maximum among all variables $j$ if one wants to compare the relative intensity among variables. We also leave an option of whether contours of the sparseness intensity are shown in the central region. 

To illustrate the visualization tools, we generate univariate functional data without outliers through a Gaussian process with zero mean and exponential covariance $C(t,s)=\exp\{-|t-s|\}$. Here we normalize the sparseness intensity by dividing it by the maximum sparseness intensity across all sparseness levels. The 1st row in Figure \ref{example} displays settings of no sparseness, medium sparseness, and high sparseness, with the corresponding sparse functional boxplots on the 2nd-row and the corresponding intensity sparse functional boxplots on the 3rd-row. Here, medium sparseness implies around 20\% missing values in every curve, whereas high sparseness implies around 60\% missing values in every curve. The magenta and gray colors in the sparse functional boxplot represent the proportion of observed values and missing values within the central region. Correspondingly, in the intensity sparse functional boxplot, the magenta color means zero sparseness intensity, the tomato color means 25\% of the maximum sparseness intensity, the gold color means 50\% of the maximum sparseness intensity, the yellow color means 75\% of the maximum sparseness intensity, and the white color means the maximum 100\% sparseness intensity.

When the data are complete, the sparse and intensity sparse functional boxplots (1st column, Figure \ref{example}) reduce to the original functional boxplot. When the sparseness level increases, the sparse functional boxplot (2nd row, Figure \ref{example}) changes from no gray area to almost a $20\%$ gray area, and then nearly $60\%$ gray area in proportion to the central region, reflecting the change of the missing proportion per time index. Simultaneously, we observe that the sparseness intensity (3rd row, Figure \ref{example}) within the central region generally increases with the increase of sparseness, which can be seen from the corresponding color change from the all-over magenta to tomato and magenta, and finally to mainly gold, yellow and few white. 

Given an intensity sparse functional boxplot, we can find the area with the most and least relative sparseness intensities and compare the relative sparseness intensity at any fixed time within the central region. Taking the bottom right intensity sparse functional boxplot as an example, we see the area with the time index between $5$ and $40$ and the value between $-1$ and $0.2$ is labeled in yellow, which indicates at least 75\% of the maximum sparseness intensity in this area. In contrast, the area close to the envelopes of the central region is colored with gold and tomato, which indicates 25\%-50\% of the maximum sparseness intensity. On the other hand, the central region in the bottom middle intensity sparse functional boxplot is mainly colored in magenta and tomato, which indicates the sparseness intensity is smaller than 25\% of the maximal sparseness intensity. In addition, the bottom middle intensity sparse functional boxplot displays that the fitted sparse values mainly concentrate on the tomato region.

\subsection{Sparse Two-Stage Visualization Tools}
Similar to the generalization of the two-stage functional boxplot (\citeauthor{dai2018functional} \citeyear{dai2018functional}) from the functional boxplot, the sparse two-stage functional boxplot and intensity sparse two-stage functional boxplot can be easily proposed by implementing the directional outlyingness (\citeauthor{dai2018directional} \citeyear{dai2018directional}).

Consider a $p$-variate stochastic process $\{\bm{\mathcal{X}}(t),~t \in \widetilde{\mathcal{T}}\}$ on $\mathbb{R}^p$ with at each time point $t$ cdf $F_{\bm{\mathcal{X}}(t)}$ that generates continuous paths on $C^p(\widetilde{\mathcal{T}})$. Take an arbitrary $\bm{Z} \in C^p(\widetilde{\mathcal{T}})$. Then $\bm{Z}(t)$ is a random vector on $\mathbb{R}^p$. These aforementioned authors first defined the outlyingness $\bm{O}(\bm{Z}(t), F_{\bm{\mathcal{X}}(t)})=\{1/D(\bm{Z}(t), F_{\bm{\mathcal{X}}(t)}) - 1\}\cdot \bm{v}$, where $D$ is a statistical depth function from $\mathbb{R}^p$, and $\bm{v}$ is the unit vector pointing from the median of $F_{\bm{\mathcal{X}}(t)}$ to $\bm{Z}(t)$. Then, they proposed the magnitude outlyingness and shape outlyingness:
$\bm{MO}(\bm{Z}, F_{\bm{\mathcal{X}}}) := \int\limits_{\widetilde{\mathcal{T}}} \bm{O}(\bm{Z}(t), F_{\bm{\mathcal{X}}(t)}){\textrm d}t \in \mathbb{R}^p$ and ${VO}(\bm{Z}, F_{\bm{\mathcal{X}}}) := \int\limits_{\widetilde{\mathcal{T}}} \{\bm{O}(\bm{Z}(t), F_{\bm{\mathcal{X}}(t)})-\bm{MO}(\bm{Z}, F_{\bm{\mathcal{X}}}) \}^\top \{\bm{O}(\bm{Z}(t), F_{\bm{\mathcal{X}}(t)})-\bm{MO}(\bm{Z}, F_{\bm{\mathcal{X}}}) \}{\textrm d}t \in \mathbb{R}$. Here $\bm{MO}$ and $\textit{VO}$ are used to measure the magnitude and shape outlyingness of a curve.

A robust Mahalanobis distance (RMD, \citeauthor{rousseeuw1984least} \citeyear{rousseeuw1984least}) is calculated between all $(\bm{MO}_i, VO_i)^\top$ $\in \mathbb{R}^{p+1}$ ($i=1,\ldots,N$) and its mean vector obtained from the minimal covariance determinant (\citeauthor{rousseeuw1999fast} \citeyear{rousseeuw1999fast}). Then, a curve $\bm{Z}_o$ is recognized as an outlier if its $(\bm{MO}_o, VO_o)^\top$ satisfies $RMD^2_{(\bm{MO}_o, VO_o)^\top} \geq C_{F_{\textit{RMD}},\alpha_f}$. Here, $C_{F_{\textit{RMD}}, \alpha_f}$ is the $(1-\alpha_f)$th percentile of Fisher's $F$ distribution, $F_{\textit{RMD}}$ (\citeauthor{hardin2005distribution} \citeyear{hardin2005distribution}), and $\alpha_f$ is the significance level.

With the directional outlyingness, we can improve the robustness of the sparse functional boxplot in a two-stage procedure. First, we obtain the set of outliers by applying directional outlyingness to the fitted data. Then, we apply the sparse functional boxplot procedure to the remaining curves. We add the detected outlying curves from stage one, with green color labelling the observed points, and gray color labelling the missing points. 

Overall, when combined, the sparse functional boxplot and the intensity sparse functional boxplot provide information about the proportion and intensity of sparse points within the central regions. The choice between the sparse functional boxplot and the sparse two-stage functional boxplot depends on the existence of outliers in the data.~When no outliers exist, there is no difference between the sparse functional boxplot and the sparse two-stage functional boxplot. Specifically, the simulation in Section \ref{simulation2} explores performances of the sparse functional boxplot and its two-stage form when detecting multivariate functional outliers.
\section{Simulation Study}
\label{sec4}
This section starts with an introduction of the data settings and a discussion of sparseness. Next, we address two problems in the simulations: first, we explore the choice of the best depth; second, we evaluate the outlier detection performances of the sparse functional boxplot and the sparse two-stage functional boxplot with the best depth.
\subsection{Simulation Settings}
\label{datasetting}
For simplicity, we assume $p=3$.~The complete data are defined on common time grids. Then we assign the point, peak, and partial sparseness to variables $1$-$3$ in the simulations. We set $N=100$, and the common times are from 50 equidistant points in $[0,1]$. We use $\bm{\mu}(t)=(5\sin(2\pi t), 5\cos(2\pi t), 5(t-1)^2)^\top$, $\bm{\epsilon}_i(t)\stackrel{iid}{\sim}\mathcal{N}_{3}(\bm{0}, \diag\{\sigma_1^2,\sigma_2^2,\sigma_3^2\})$. Here, $\sigma^2_1, \sigma^2_2, \sigma^2_3$ are independent and follow $U[0.3, 0.5]$, where $U$ means a random sampling from the uniform distribution.
We use orthonormal Fourier basis functions to construct ${\psi}^{(j)}_m(t)$. We let $M=9$, $\rho_{i,m}\stackrel{iid}{\sim}\mathcal{N}(0,\nu_m)$, and $\nu_m=\frac{M+1-m}{M}$.

Eight models are provided by specifying the eigenfunctions and outliers as below. Model 1 is a reference model without contamination, while the remaining models include contamination by adjusting $\bm{\mu}(t)$, $\bm{u}_{i}(t)$, or $\bm{\epsilon}_i(t)$, with a contamination level of 10\%. Models 2-4 introduce magnitude outliers (\citeauthor{sun2011functional} \citeyear{sun2011functional}), and Models 5-8 generate shape outliers (\citeauthor{dai2018multivariate} \citeyear{dai2018multivariate}, \citeauthor{dai2018directional} \citeyear{dai2018directional}).

\textbf{Model~1} (no outlier): $\footnotesize{
    {\bm{Y}}_i(t)=\bm{\mu}(t)+\bm{u}_{i}(t)+\bm{\epsilon}_i(t)= \bm{\mu}(t)+\sum_{m=1}^{M}\rho_{i,m}\bm{\psi}_m(t)+\bm{\epsilon}_i(t), ~~~i=1,\ldots,N.
    }$
    
\textbf{Model~2} (persistent magnitude outliers):
$u^{(j)}_{i,ou}(t)=
 u^{(j)}_{i}(t)+8w_{i,j}~{\rm for}~t\in [0, 1]$, where $w_{i,j}~(j=1,2,3)$ follows the binomial distribution of obtaining $1$ with probability $0.5$, and $-1$ with probability $0.5$ and all $w_{i,j}$s in the manuscript follow the same distribution as in Model~2.

\textbf{Model~3} (isolated magnitude outliers): $u^{(j)}_{i,ou}(t)=u^{(j)}_{i}(t)+8w_{i,j}~ {\rm for}~t \in [T_{s},T_{s}+0.1]$, and $T_{s}$ is from $U[0,0.9]$.

\textbf{Model~4} (shape outliers I): $\bm{\mu}_{ou}(t)=(\mu^{(1)}(t-0.3),\mu^{(2)}(t-0.2),\mu^{(3)}(t-0.5))^\top$.

\textbf{Model~5} (shape outliers II):
\begin{small}
    $u^{(1)}_{i,ou}(t)=u^{(1)}_{i}(t)+2\sin(4\pi t),
 u^{(2)}_{i,ou}(t)=u^{(2)}_{i}(t)+2\cos(4\pi t),$
$u^{(3)}_{i,ou}(t)=u^{(3)}_{i}(t)+2\cos(8\pi t),
    u^{(j)}_{i,no}(t)=u^{(j)}_{i}(t)+U_j,~j=1,2,3,$
\end{small}where $U_j$ is generated from $U[-2.1,2.1]$.

\textbf{Model~6} (mixed outliers):
\begin{small}$
\bm{\mu}_{i,ou}(t)=((2+R^{(1)}_i)\mu^{(1)}(t),(2+R^{(2)}_i)\mu^{(2)}(t),(2+R^{(3)}_i)\mu^{(3)}(t)-6)^\top,~j=1,2,3,$
\end{small} where $R^{(j)}_i$ follows ${\rm Exp}(2)$, and ${\rm Exp}$ is a random sampling from the exponential distribution.

\textbf{Model~7} (joint outliers):
\begin{small}$u^{(1)}_{i,ou}(t)= Z_1 t\sin(\pi t),~
 u^{(2)}_{i,ou}(t)=Z_2 t\cos(\pi t),~
 u^{(3)}_{i,ou}(t)= Z_3 t\sin(2\pi t),$
  $ u^{(1)}_{i,no}(t)=Z_4 t\sin(\pi t),~
 u^{(2)}_{i,no}(t)= (8-Z_4) t\cos(\pi t),~
 u^{(3)}_{i,no}(t)= (Z_4-2) t\sin(2\pi t),$
\end{small}
where $Z_1$-$Z_4$ are generated from $U[2,8]$.

\textbf{Model~8} (covariance outliers):
$\bm{\epsilon}_i(t)$ are generated from a stationary isotropic cross-covariance model belonging to the Mat\'ern family (\citeauthor{gneiting2010matern} \citeyear{gneiting2010matern}) with smoothness parameters $\nu_{ij} > 0$: $C_{ii}(s,t)=\sigma^2_i \mathcal{M}(|s-t|;\nu_{ii})$ for $i=1,2,3,$ and $C_{ij}(s,t)=\rho_{ij}\sigma_i\sigma_j \mathcal{M}(|s-t|;\nu_{ij})$ for $1\leq i \neq j\leq 3,$ and $\mathcal{M}(r;\nu_{ij}) = \frac{2^{1- \nu_{ij}}}{\Gamma(\nu_{ij})} (\sqrt{2\nu_{ij}} r)^{\nu_{ij}} K_{\nu_{ij}}(\sqrt{2\nu_{ij}} r)$, where $K_{\nu}$ is the modified Bessel function of the second kind of order $\nu$, $\nu_{ii}$ is generated from the uniform distribution, with ranges $[2,3]$ for the non-outliers and [0.1,0.2] for the outliers, $\nu_{ij}=\frac{\nu_{ii}+\nu_{jj}}{2}$, $\rho_{ii}=1$, and $\rho_{ij} = \frac{i+j}{i+j+3}$.

To categorize possible sparseness scenarios, we consider three sparseness types in marginal functional data: point, peak, and partial sparseness. Point sparseness means that missing points appear randomly in time grids, and the location and number of missing points are independent per curve. The peak and partial sparseness mean that values are missing during a continuous interval starting from a random point $t_{start}$. The difference between peak and partial sparseness lies in the $t_{start}$. In the peak sparseness, for each curve, $t_{start}$ is generated independently, while in the partial sparseness, all curves with missing values share a common $t_{start}$. Define $p_{size}$ (the number of curves with missing values divided by the total number of curves) as the probability of a curve with at least one missing value in all sample curves, and $p_{curve}$ (the number of missing points in the curve divided by the number of complete-time measurements in the curve) as the probability of sparseness in each of those curves with missing values. Hence, we use the sparseness parameter $\bm{p}_s=(p_{size}, p_{curve}, t_{start})^\top$ to tune the sparseness. 
\begin{figure}[h!]
   \centering
   \includegraphics[width=1\textwidth,height=5cm]{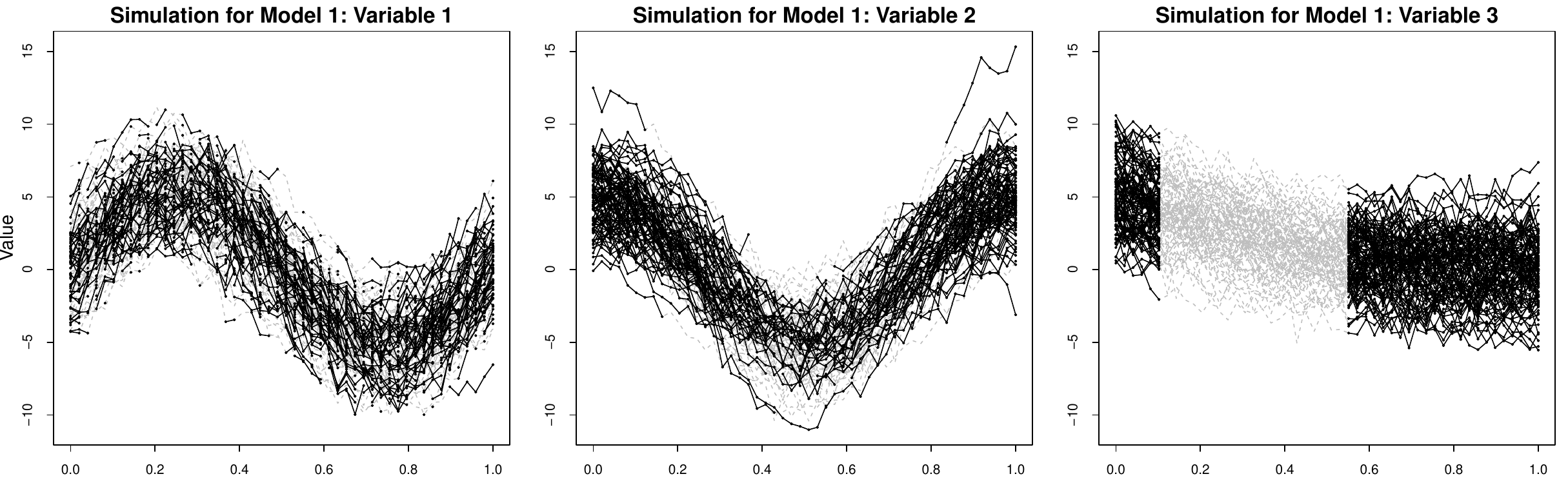}
   \includegraphics[width=1\textwidth,height=5cm]{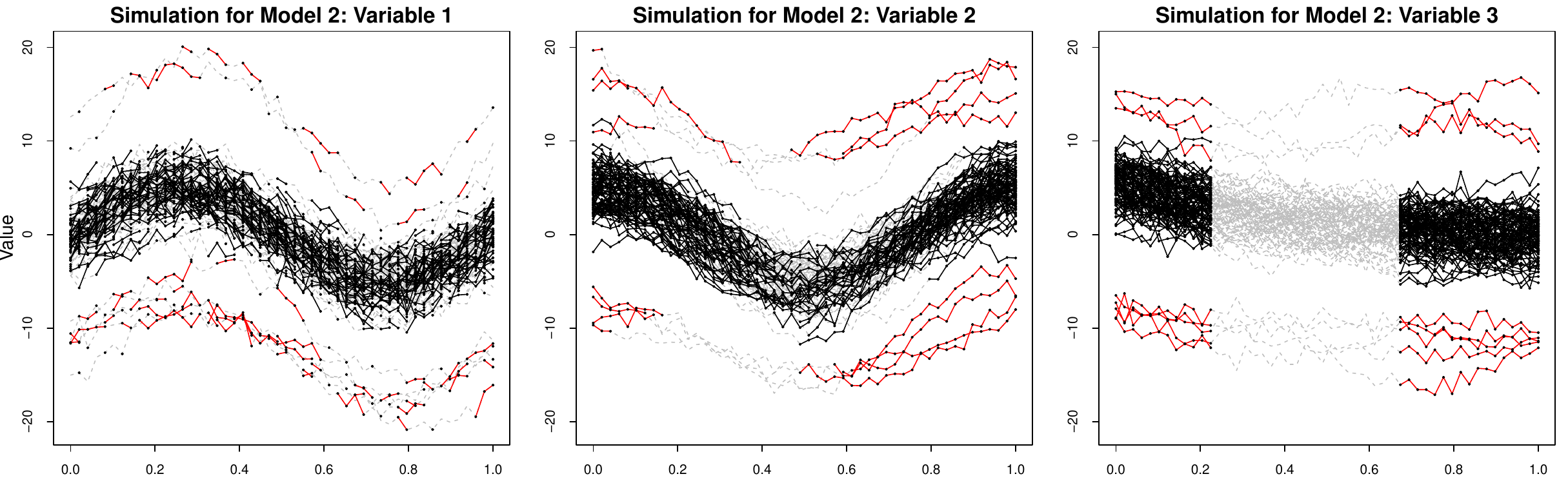}
   \includegraphics[width=1\textwidth,height=5cm]{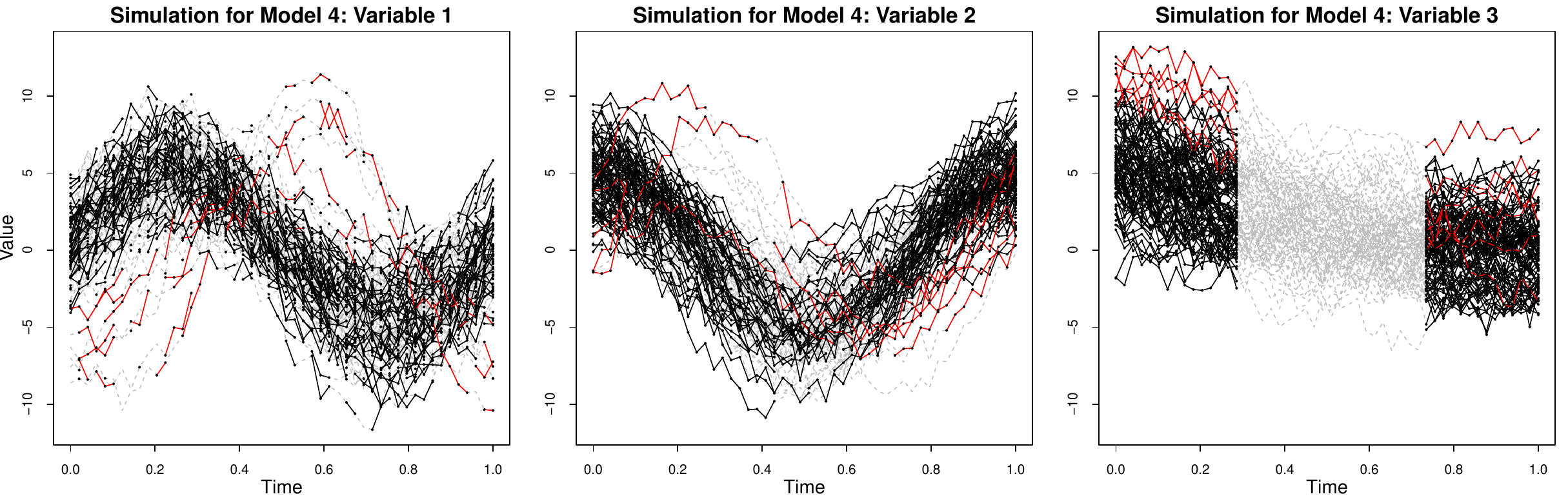}
   \caption{Simulations for Models 1 (no outlier), 2 (persistent magnitude outliers), and 4 (shape outliers I), with $p_{curve}=40\%$ and $p_{size}=100\%$. Typical curves are colored in black in the simulation plot, and outliers are colored in red, with the observed points shown as black dots. Artificial sparseness is represented by gray dashed lines. We assign the point, peak, and partial sparseness to variables $1$, $2$, and $3$ in the simulation.}
    \label{sim_org}
\end{figure}

It is worth noting that $t_{start}$ is only required in the peak and partial sparseness as a starting point of the missing interval. Here, $t_{start}$ is a realization from $U[0,1-p_{curve}]$. Usually, the missing interval does not cover the first or the last time point in $[0, 1]$. Otherwise, it is difficult for BMFPCA to reconstruct curves with suitable fittings unless the curves follow simple monotone trends, such as for the data application in Section \ref{app2}.

We display a simulation for Models 1, 2, 4 (Figure \ref{sim_org}) to illustrate the sparseness types and $\bm{p}_s$, while the remaining simulation visualizations are shown in the Supplementary Material. In Figure \ref{sim_org}, variables 1-3 belong to the point (1st column), peak (2nd column), and partial (3rd column) sparseness types, respectively with $N = 100$ and contamination level 10\%. Here, $p_{size} = 100\%$ means that 100 curves from the simulation contain missing values, and $p_{curve} = 40\%$ means that in each curve with missing values, there are about 20 time measurements with missing values given 50 time measurements. We can see that each curve has different missing time points per curve in the point sparseness (1st column, Figure \ref{sim_org}). In addition, we also observe a missing interval with the various start of missing points per curve (2nd column, Figure \ref{sim_org}) in the peak sparseness, and a missing interval with a common start of missing points  (3rd column, Figure \ref{sim_org}) in the partial sparseness.

 \subsection{Simulation I: Choice of Depth}
 \label{simulation1}
We apply the conventional depth to the MFPCA and BMFPCA fits (Eq. (\ref{model-based-estimate})-(\ref{eq})), and name them MFHD$_{mfpca}$ and MFHD$_{bmfpca}$, respectively. We set $B = 100$ in the BMFPCA fit. The BMFPCA fit is more robust and closer to the original data than the MFPCA fit. Hence, we apply the remaining revised depths RMFHD$_{aw}$ and RMFHD$_{naw}$ (Eq. (\ref{md1})) to the BMFPCA fit.  
\begin{figure}[b!]
    \centering
   \includegraphics[width=1\textwidth,height=5.55cm]{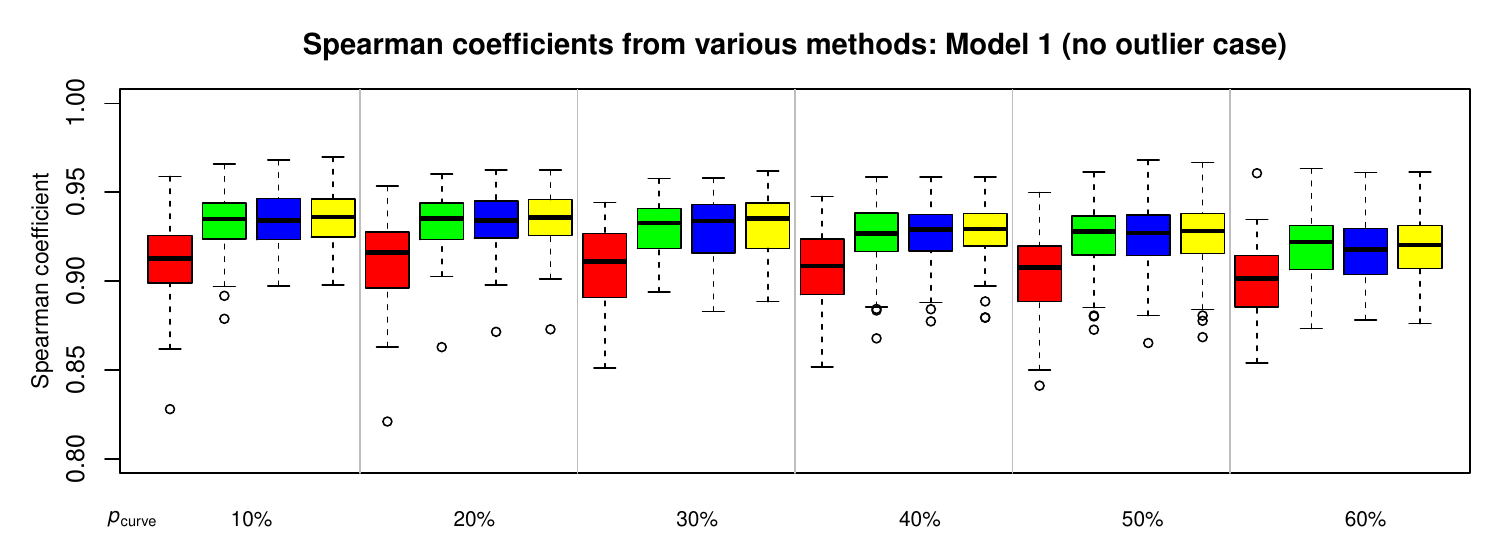}
\includegraphics[width=1\textwidth,height=5.55cm]{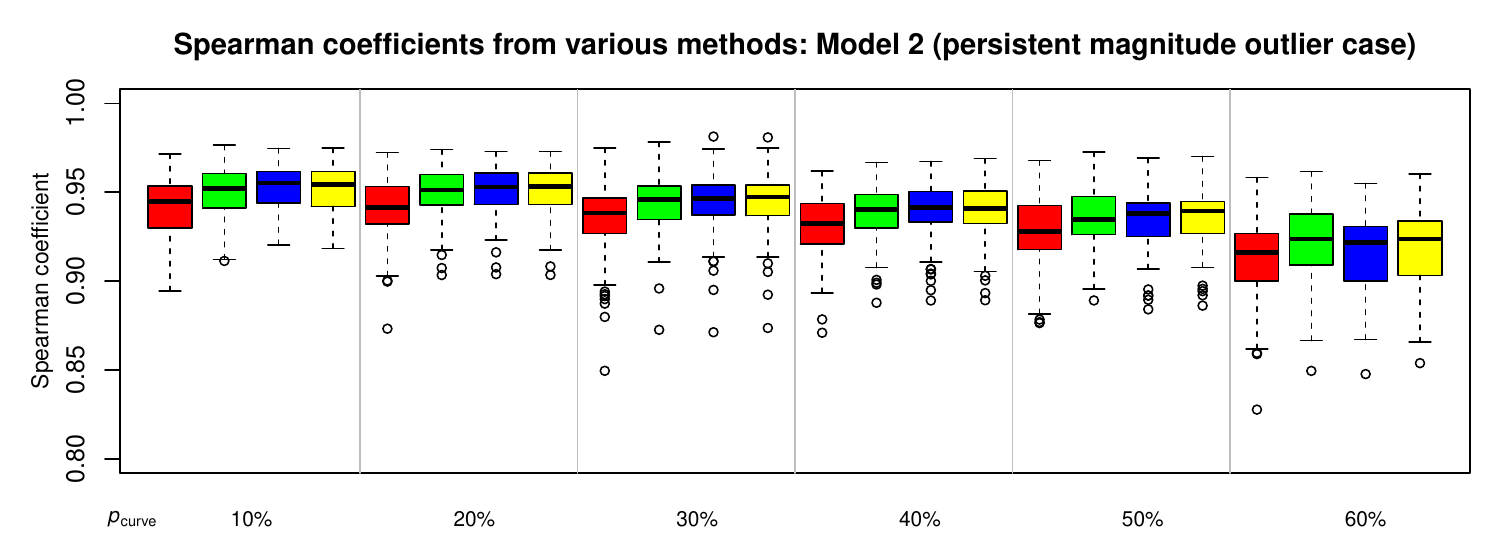}
\includegraphics[width=1\textwidth,height=6.05cm]{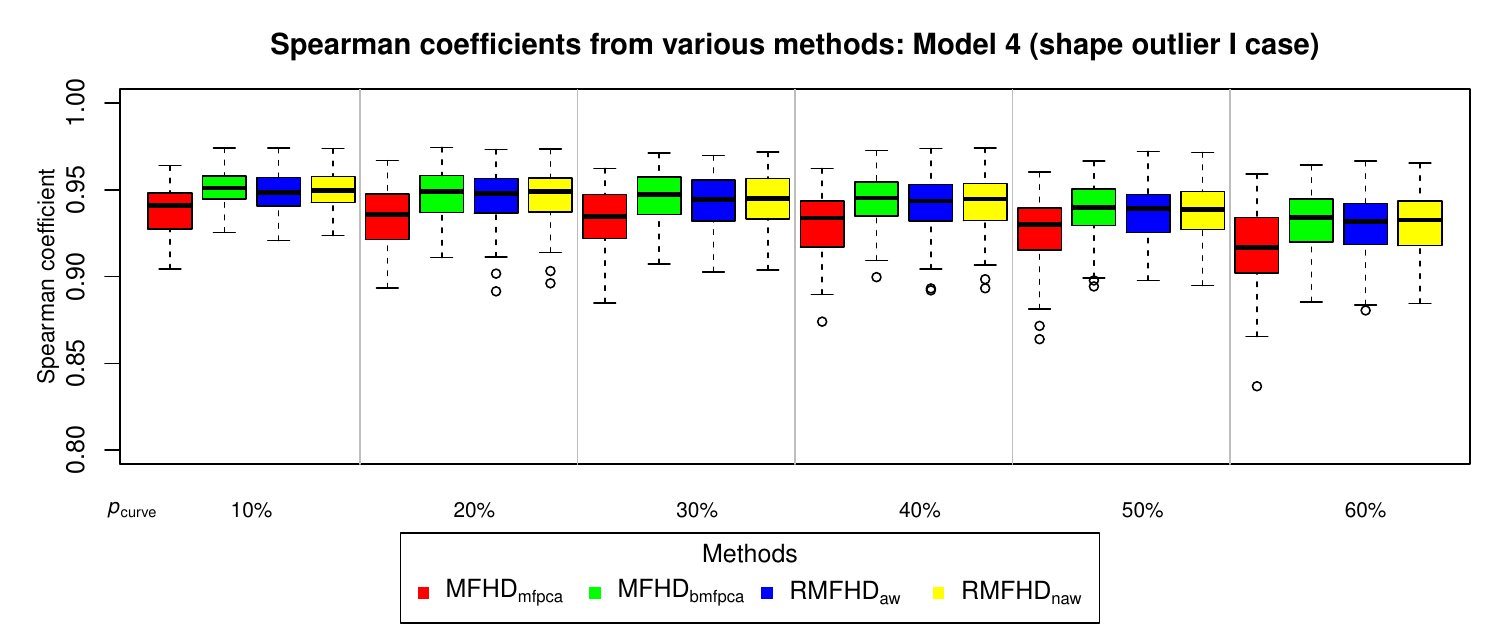}
\caption{Comparison of four depth ranking methods for Models 1, 2 and 4. The panels from top to bottom represent boxplots of Spearman coefficients under various data settings. In each setting, $p_{curve}$ is given from $10\%$ to $60\%$, with $p_{size}=100\%$ in the \textbf{point} \textbf{sparseness} type. The number of replicates is 100.}
\label{spearman}
\end{figure}

The Spearman rank correlation coefficient (\citeauthor{kendall1945treatment} \citeyear{kendall1945treatment}) assesses how well the rank based on the fitted data, given by the revised depth, correctly associates with the rank based on the original complete data, given by the conventional depth. When the coefficient is closer to 1, the stronger the association, and when it is closer to 0, the weaker the association.

 The Spearman coefficients of the above four methods are presented in Figure \ref{spearman} under various $p_{curve}$ with $p_{size}=100\%$ for Models 1, 2, and 4 when the point sparseness exists in all variables. The results of the remaining models from the point sparseness and those from other sparseness types can be seen in the Supplementary Material. All depths show a bit of weaker association when the curve is sparser on average in the peak and partial sparseness types. Generally, the application of MFHD to the BMFPCA has a stronger rank association than the application of MFHD to the MFPCA with the original data and have a slight advantage over the revised depth $MFHD_{aw}$ and $MFHD_{naw}$.

Overall, MFHD$_{bmfpca}$ provides excellent performance with its strong rank associated with the original data and simple procedure of application for sparse multivariate functional data. It omits the confidence band construction and thus, requires no additional depth modifications and is more efficient in computation.
\subsection{Simulation II: Choice of Visualization Tools}
\label{simulation2}
We apply the sparse functional boxplot and the sparse two-stage functional boxplot in different scenarios with MFHD$_{bmfpca}$.~The two-stage functional boxplot detects both magnitude and shape outliers with the help of directional outlyingness. We use $p_c$, the correct detection rate (the number of correctly detected outliers divided by the number of outliers) and $p_f$, the false detection rate (the number of falsely detected outliers divided by the number of non-outliers) to measure the performance of the above two tools in multivariate functional outlier detection. 
\begin{table}[h!]
\begin{center}
\caption{The mean and standard deviation (in parenthesis) of the percentages $p_c$ and $p_f$ for the sparse functional boxplot and the sparse two-stage functional boxplot with 100 replications and 100 curves when the curve sparseness $p_{curve}$ is 20\%, 40\%, and 60\%, respectively, and $p_{size}=100\%$ for the point sparseness type, in the presence of Models 2 (persistent magnitude outliers) and 4 (shape outliers I). The probability of outliers in each model is 10\%.}
\label{tab}
\begin{small}
\begin{tabular}{ c|c|c|c|c|c|c } 
 \hline
 \hline
 Curve Sparseness & \multicolumn{2}{c|}{$p_{curve}=20\%$}&  \multicolumn{2}{c|} {$p_{curve}=40\%$}& \multicolumn{2}{c}{$p_{curve}=60\%$}  \\
  \hline
 \diagbox{Methods}{Ratios} &  ${p}_c$ & ${p}_f$ & ${p}_c$ & ${p}_f$ & ${p}_c$ & ${p}_f$\\  
   \hline
 Sparse: Model 2 & 64.9 (25.0) & 0.0 (0.0) &61.4 (23.5) & 0.0 (0.0) & 66.0 (25.5)& 0.0 (0.0) \\ 
 \hline
 Sparse: Model 4 & 15.7 (19.4) & 0.0 (0.0)  &15.4 (21.4) & 0.0 (0.0) &13.5 (19.4) & 0.0 (0.0) \\ 
 \hline
 Sparse Two-Stage: Model 2 & 100.0 (0.0) &  0.0 (0.0) &100.0 (0.0) & 0.0 (0.0)& 100.0 (0.0) & 0.0 (0.0) \\ 
  \hline
 Sparse Two-Stage: Model 4 &97.8 (4.9) & 0.1 (0.5) &97.1 (6.3) & 0.1 (0.4)& 97.5 (4.8) & 0.1 (0.4)  \\ 
 \hline
 \end{tabular}
 \label{outl_table}
 \end{small}
\end{center}
\vspace{-0.5cm}
\end{table}

We display in Table \ref{outl_table} the performance of $p_c$ and $p_f$ for Models 2 and 4 when the point sparseness is assigned to all variables, while the other performances are provided in the Supplementary Material. Overall, the sparse two-stage functional boxplot performs better in detecting true outliers than the sparse functional boxplot, at the expense of a slightly higher $p_f$.

BMFPCA fits the curves after removing the mean trend; hence, it manages to capture the typical behavior of the curves independently of their amplitude. Suppose outliers show an abnormality, especially in the shifted amplitude or time. In that case, the sparse two-stage functional boxplot obtains a high $p_c$, see Models 2 (persistent magnitude outliers) and 4 (shape outliers I). We see a somewhat smaller $p_c$, between 50\% and 80\%, from the sparse two-stage functional boxplot in Models 3 (isolated magnitude outliers), 6 (mixed outliers), and 7 (joint outliers), where the abnormality mainly appears in an interval of the outliers. Furthermore, we notice that the sparse two-stage functional boxplot does not detect outliers well in Models 5 (shape outliers II) and 8 (covariance outliers), showing a smoothness difference between abnormal and normal curves. It is difficult to detect the outlying samples due to not many smoothness differences among the fitted curves. On the whole, we recommend the sparse two-stage functional boxplot for detecting potential outliers.

\section{Applications}
\label{sec5}
Applications include the univariate CD4 cell counts and bivariate malnutrition data introduced in Section \ref{sec:intro}. The analysis follows the procedures of fitting the data and consideration between the sparse functional boxplot and its two-stage form. We do not show contours in the intensity sparse two-stage functional boxplot for the CD4 cell counts due to the narrow range in the central region, but we offer them for the malnutrition data. 
\subsection{Univariate Case: CD4 Cell Counts Data}
The human immune deficiency virus (HIV) harms the body by attacking an immune cell called the CD4 cell and, thus, making the body more vulnerable to other illness-causing germs. Hence, CD4 cell count per milliliter of blood (\citeauthor{taylor1989cd4} \citeyear{taylor1989cd4}) can be used to track the progression of HIV. A person with untreated HIV will experience a series of stages: acute HIV infection (2-6 weeks), stage one (1-5 years), stage two (6-9 years), stage three (9-11 years), which is advanced HIV disease, and stage four (11-12 years), which is called the acquired immunodeficiency syndrome (AIDS). In the acute HIV infection and in stage one, the CD4 cell count drops slowly and is usually above 500; during stage two, it ranges between 350 and 499. In the stage three, the CD4 count is 200 to 349, whereas in the stage four the CD4 count is less than 200 (\citeauthor{hivg} \citeyear{hivg}). These CD4 cell counts data are available in the \textit{refund} package (\citeauthor{crainiceanu2010bayesian} \citeyear{crainiceanu2010bayesian}) in R and were analysed by \citeauthor{goldsmith2013corrected} (\citeyear{goldsmith2013corrected}) to construct the bootstrap improved confidence bands for sparse functional data. It includes observed CD4 cell counts for 366 infected individuals from $-18$ to $42$ months (Figure \ref{cd4}), where 0 represents seroconversion, the moment when the antibody becomes detectable in the blood. The observation period in this data set includes acute HIV infection and stage one, or stage two for some patients who are in worse situations. 

\begin{figure}[!t]
   \begin{minipage}{0.49\textwidth}
     \centering
     \includegraphics[width = .95\linewidth, height = 7cm]{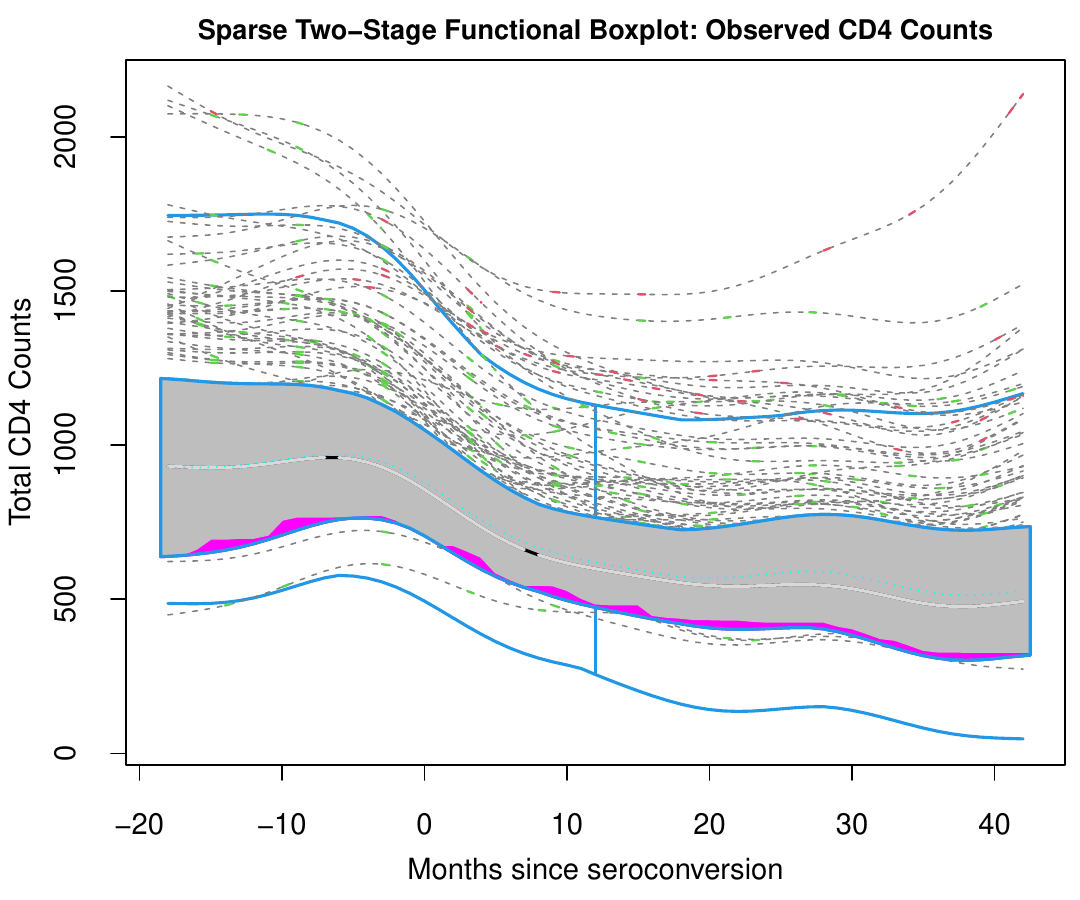}
   \end{minipage}\hfill
   \begin{minipage}{0.51\textwidth}
     \centering
     \includegraphics[width = 1\linewidth, height = 7cm]{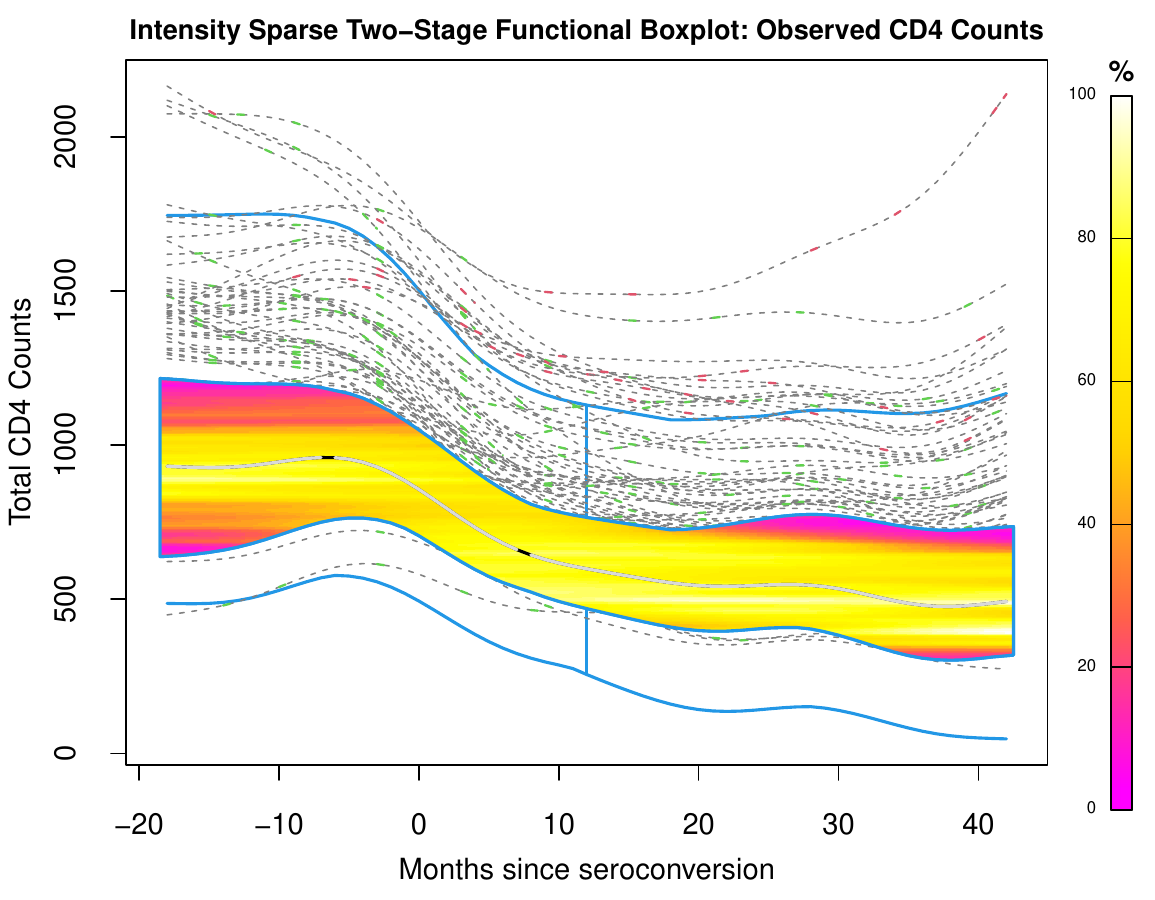}
   \end{minipage}
   \caption{The sparse two-stage functional boxplot and the intensity sparse two-stage functional boxplot for observed CD4 cell counts.~Seventy-eight outliers (green dashes) are detected from directional outlyingness, and eight outliers (red dashes) are detected from the functional boxplot.}
   \label{visualization_cd4}
\end{figure}

 Figure \ref{visualization_cd4} contains the CD4 cell counts after they were fitted with the iterated expectation from UFPCA, and ordered according to the modified band depth (MBD, \citeauthor{lopez2009concept} \citeyear{lopez2009concept}). Compared to the original sparse observations shown in Figure \ref{cd4}, the fit ranges between 20 and 2173, which lies in the field of low and normal CD4 cell count per cubic millimeter of blood. The CD4 cell count is a point sparseness case with $p_{size} = 100.0\%$ and $p_{curve}$ is between 82.0\% (11 observations per subject) and 98.4\% (1 observation), which is displayed by the magenta area in the sparse two-stage functional boxplot in Figure \ref{visualization_cd4}. The sparse functional boxplot detects seven outliers with a contamination level of 3.0\%. The sparse two-stage functional boxplot detects 78 outliers (green dashed lines) from directional outlyingness and eight extra outliers (red dashed lines) from the functional boxplot, with a contamination level of 23.0\%. Additionally, the intensity sparse two-stage functional boxplot indicates that gold and yellow colors fill most central parts within the central region, which correspond to 50-75\% of the maximal sparseness intensity. Conversely, the region with small sparseness intensity mainly lies in the magenta areas, one at the start when the CD4 estimated counts are below 700 and above 1200, and another between 20-42 months after seroconversion when the CD4 cell counts are below 400 and above 800.

 The 50\% central region starts between 631 and 1497 at 18 months before seroconversion and ends between 316 and 744 at 42 months after seroconversion.~The median shows a change in CD4 cell counts from 935 to 500 during the observation period. There are 27.5\% of patients who show CD4 cell counts below 350 at 42 months after seroconversion, implying that they are close to, or already at, stage three. However, 89.5\% of outliers show a rising trend of CD4 cell counts above 500 in the period after seroconversion, suggesting that they are on the right track against HIV. 
 
 In this analysis, we wish to review the progression of an HIV infection during different observation periods according to the visualization tools we introduced. In the window period, before the HIV antibody can be detected in the blood, CD4 cell counts have an inverse U shape, which can be seen from outliers and the upper fence of the central region (Figure \ref{visualization_cd4}) between 10 months before and 10 months after the seroconversion, implying the process of rapid replication and severe drop in CD4 cells. This reflects the HIV first invading the body, stimulating CD4 cells to replicate, and then infecting CD4 cells by injecting them with its genetic material, leading to a fast drop in the number of CD4 cells. Surprisingly, there are 21 patients with CD4 cell counts around or above 1000 at 42 months after seroconversion, whose treatment or lifestyle may deserve further study.
\subsection{Bivariate Case: Malnutrition Data}
\label{app2}
The malnutrition data, from the United Nations Children's Fund (\citeauthor{malnutrition} \citeyear{malnutrition}) data warehouse, include two variables, stunted growth and the prevalence of low birth weight, collected in 77 countries from 1985 to 2019. Stunted growth is defined as the proportion of newborns aging from 0 to 59 months with a low height-for-age measurement (below two standard deviations). The stunted growth data represent a point sparseness case with 4-23 recordings per nation; $p_{curve}$ ranges from $29.4\%$ in Bangladesh to $88.2\%$ in Argentina. The low birth weight data are a partial sparseness case, with recordings during 2000-2015 only; $p_{curve}=52.9\%$ for all nations. 
\begin{figure}[b!]
  \centering
  \begin{flushleft}
  \includegraphics[width=0.93\textwidth,height=7cm]{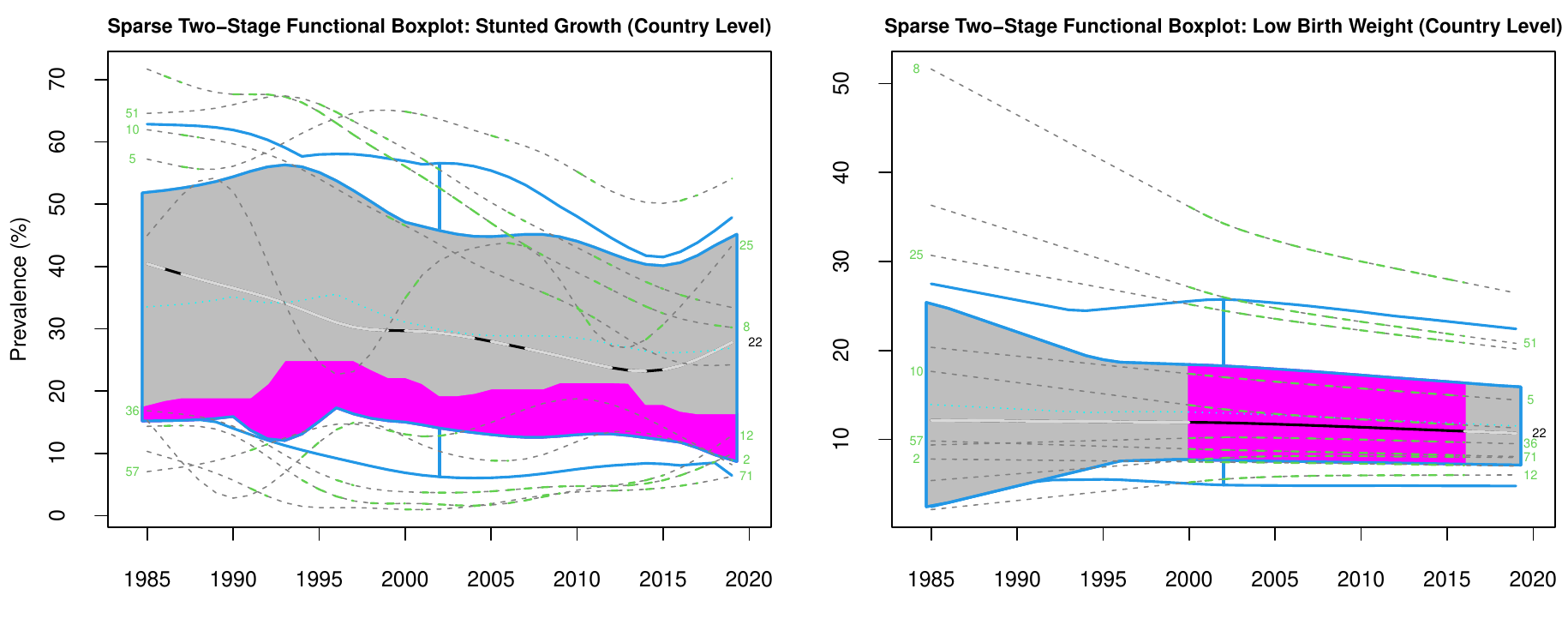}
  \includegraphics[width=0.97\textwidth,height=7cm]{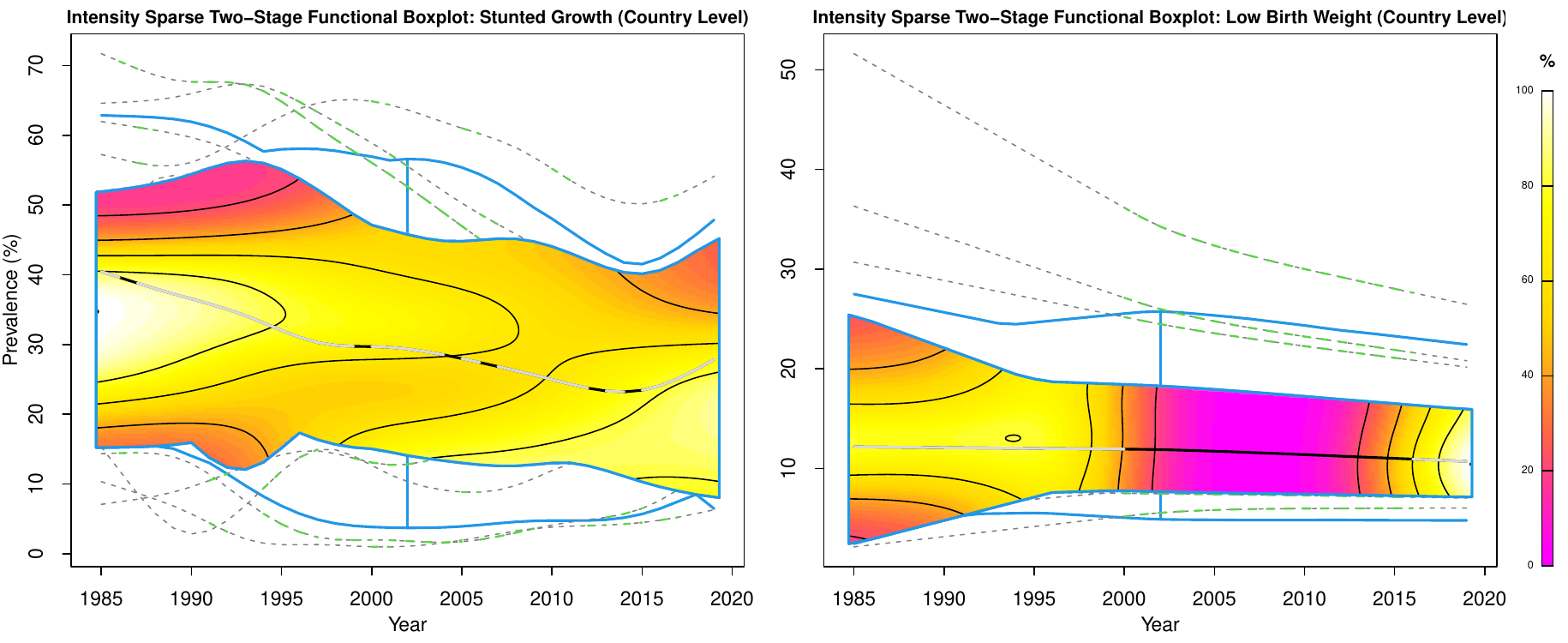}
  \end{flushleft}
  \vspace{-0.5cm}
 \caption{Visualization of stunted growth and low birth weight data for 77 countries with the sparse two-stage functional boxplot and the intensity sparse two-stage functional boxplot. One-to-one maps (number, outlier/median nation) are the following: (2, Argentina), (5, Burundi), (8, Bangladesh), (10, Bhutan), (12, Chile), (22, Ecuador), (25, Guinea-Bissau), (36, Kuwait), (51, Nepal), (57, Romania), and (71, United States).}
 \label{mal}
\end{figure}

 Since only one outlier (Bangladesh) is detected as a magnitude outlier in the prevalence of the low birth weight via the sparse functional boxplot, we implemented the sparse two-stage functional boxplot and detected ten outliers. First, the grey and magenta in the central region (top row of Figure \ref{mal}) display the sparseness proportion of the central region for each variable. In addition, the median and fences of the central region display that most nations went through a significant decline in stunted growth prevalence over time, while most countries had a slow drop or rise before 2000 and a steady trend since 2000 in low birth weight. Furthermore, the low birth weight prevalence is right-skewed which can be seen from the margin difference between the central region and the fences. The skewed setting suggests more attention to nations with high prevalence of low birth weight is needed.
 
The intensity sparse two-stage functional boxplot (bottom row, Figure \ref{mal}) displays the relative intensity of the fitted missing values inside the central region for each variable. The missing values for the prevalence of low birth weight are most intense in the almost white region: with the interval 1985-1988 and the estimates around 30\%-40\%. Comparatively, the missing values with the most intensity for the low birth weight lie in the region where the estimates are between 8\% and 12\% and time is between 2018-2019. In addition, we can see the relative intensity of fitted missing values in both variables at each fixed time. For instance, for the stunted growth, when the year is 1985, missing values are mostly distributed between 25\%-40\% prevalence labeled in white, followed by the prevalence between 21\%-25\% and 40\%-42\% labeled in yellow, after that the prevalence between 19-21\% and 42-44\% labeled in gold, then the prevalence between 16\%-19\% and 45\%-49\% labeled in tomato, succeeded by the prevalence above 49\% including the least sparseness intensity labeled in magenta.

We use the human development index (\citeauthor{undp} \citeyear{undp}), which categorizes countries into low, medium, high, and very high human development groups, to analyze the malnutrition trend in these countries. In the last 35 years, the median, Ecuador, shows a representative trend of a smooth decline from 40\% to 24\% in the prevalence of stunted growth, and a constant trend around 12\% in the prevalence of low birth weight. Outliers exist in almost all human development groups, often showing an abnormality in the shape of stunted growth prevalence. From the very high human development group, Argentina, Chile, Kuwait, Romania, and the United States demonstrate an exception in shape lying at the bottom of the stunted growth's prevalence and a slow rise at the bottom of the low birth weight's prevalence. From the middle human development group, Bhutan shows only an abnormal shape in the prevalence of stunted growth, which is not far from their respective median. Finally, Bangladesh, Burundi, Guinea-Bissau, and Nepal, which all belong to the low human development index group, lie at the top of the samples in Figure~\ref{mal}. They all show a faster decline rate in the prevalence of the stunted growth compared to the rest of their group and a significant decline, especially for Bangladesh and Nepal, in the prevalence of the low birth weight. Nevertheless, they are on top of the curves in 2019 and need more actions to reduce malnutrition.

\section{Discussion}
\label{sec6}
In this paper, we proposed two flexible tools, the sparse functional boxplot and the intensity sparse functional boxplot, for visualizing sparse multivariate functional data. We also introduced a sparse form of the two-stage functional boxplot (\citeauthor{dai2018functional} \citeyear{dai2018functional}), which is itself a derivation from the functional boxplot (\citeauthor{sun2011functional} \citeyear{sun2011functional}) with better outlier detection. All of these can be applied in both univariate and multivariate functional settings. When the data are observed on common time grids without missing values, the tools reduce to the original functional boxplot and two-stage functional boxplot. We believe the introduction of visualization tools for sparse data is of great importance due to the wide applications of longitudinal studies and the common challenge of missing values in real data sets. 

To apply the visualization tools described above to sparse functional data, an appropriate data fitting and depth for data ordering is required. In addition to the novel contribution in the sparse and intensity sparse functional boxplot, we improved the fitting of data (MFPCA, \citeauthor{happ2018multivariate} \citeyear{happ2018multivariate}) through the iterated expectation from bootstrap (\citeauthor{goldsmith2013corrected} \citeyear{goldsmith2013corrected}), and investigated several depths for sparse multivariate functional data. Here, we took the multivariate functional halfspace depth (MFHD, \citeauthor{claeskens2014multivariate} \citeyear{claeskens2014multivariate}) as a building block for proposing various revised depths for sparse multivariate functional data. We obtained the best depth via the Spearman rank coefficient simulated in various data settings and sparseness scenarios. In the univariate functional setting, MFPCA became UFPCA (\citeauthor{yao2005functional} \citeyear{yao2005functional}), and we used the modified band depth (MBD, \citeauthor{lopez2009concept} \citeyear{lopez2009concept}) to order the data, prior to the application of visualization tools.

 Besides data visualization, the sparse functional boxplot and its two-stage form can also detect outliers. Simulations demonstrated that generally, the sparse two-stage functional boxplot performed better than the sparse functional boxplot in outlier detection. In some shape outlier cases, the advantage of outlier detection from directional outlyingness (\citeauthor{dai2018directional} \citeyear{dai2018directional}) is not obvious, and may require further innovations in outlier detection. Two public health applications, CD4 cell counts in individuals and malnutrition data at national levels, displayed how information may be extracted from the sparse two-stage functional boxplot and the intensity sparse two-stage functional boxplot.~Extensions of the methods proposed in this paper to sparse images and surfaces could be explored with the surface boxplot (\citeauthor{genton2014surface} \citeyear{genton2014surface}).   

\begin{center}
  \section* {SUPPLEMENTARY MATERIALS}  
\end{center}
\textbf{Supplements:} \textit{R-code for sparse and intensity functional boxplots:} R-code for the commands sparse\_fbplot and intensity\_sparse\_fbplot described in the articles (sparse\_fbplot.R \& intensity\_sparse\_fbplot.R).
 \textit{Simulation code:} Simulation code for the optimal depth under eight models described in the article (00\_execute\_simulation\_spearman.R) and the outlier detection under eight models described in the article (execute\_outldetect.R). \textit{CD4 data:} CD4 count for 366 subjects from 18 months before to 42 months after seroconversion, load cd4 (\citeauthor{goldsmith2013corrected} \citeyear{goldsmith2013corrected}) in $refund$ package.
 \textit{Malnutrition data:} The original sparse prevalence of stunted growth and low birth height from 1985 to 2019 (malnutrition.csv). \textit{Supplementary Material}: The performances of several depths and outlier detection under eight models in different sparseness types. All files can be found in a single zip file (sparse\_fbplot.zip).
~\\

\setlength{\bibsep}{5pt}
\bibliographystyle{apalike2}
\bibliography{ref.bib}

\end{document}